\documentstyle[aps,prb,epsfig,floats]{revtex}

\newcommand{\bi}{{-}}
\newcommand{\blamk}{{\bf k}}
\newcommand{\bnul}{{\bf 0}}
\newcommand{\C}{{\bar{\bar{\mbox{$C$}}}}}
\newcommand{\bC}{{\bar{\bar{\mbox{$C$}}}}}
\newcommand{\bD}{{\bar{\bar{ \mbox{$D$}}}}}
\newcommand{\drl}{{\Delta r_{||}}}
\newcommand{\bn}{{\mathbf{n}}}
\newcommand{\coulener}{{\varepsilon_{\mathrm{C}}}}
\newcommand{\drt}{{\Delta r_{\perp}}}
\newcommand{\ekstra}[1]{#1}
\newcommand{\F}{{\bar{\bar{\mbox{$F$}}}}}
\newcommand{\bF}{{\bar{\bar{ \mbox{$F$}}}}}
\newcommand{\fin}{{+}}
\newcommand{\g}{g}
\newcommand{\gc}{\g_1}
\newcommand{\gbl}{\langle \hspace{- 1 mm}  \langle \hspace{ - 1 mm}\langle}
\newcommand{\gbr}{\rangle \hspace{- 1 mm}  \rangle \hspace{ - 1 mm}\rangle}
\newcommand{\matrixcomma}{\,,}
\newcommand{\pderivet}{{\partial_{t}}}
\newcommand{\pderives}{{\partial_{s}}}
\newcommand{\permittivity}{{\epsilon}}

\renewcommand{\Re}{{\text{Re}}}
\newcommand{\R}{{\bar{\bar{\mbox{$R$}}}}}
\newcommand{\Sint}{{S_{\text{int}}}}
\newcommand{\zhat}{{\hat{\bf z}}}
\newcommand{\unitz}{{\zhat}}
\renewcommand{\Im}{{\text{Im}}}

\newcommand{\avell}{\langle \hspace{- 0.7 mm}  |}
\newcommand{\averl}{| \hspace{- 0.7 mm}\rangle  }
\newcommand{\avelb}{\big{\langle} \hspace{- 1 mm} \big{ |} }
\newcommand{\averb}{\big{|} \hspace{- 0.9 mm}  \big{\rangle} }

\newcommand{\avelBb}{\Bigg{\langle} \hspace{- 1.05 mm} \Bigg{ |} }
\newcommand{\averBb}{\Bigg{|} \hspace{- 1 mm}  \Bigg{\rangle} }

\newcommand{\pilf}{\pill }
\newcommand{\pirf}{\pirl }
\newcommand{\pill}{\langle \hspace{- 1 mm}  \langle }
\newcommand{\pirl}{\rangle \hspace{- 1 mm}  \rangle }
\newcommand{\pilB}{\bigg{\langle} \hspace{- 2.2 mm}  \bigg{\langle} }
\newcommand{\pirB}{\bigg{\rangle} \hspace{- 2.2 mm}  \bigg{\rangle} }
\newcommand{\pilBb}{\Bigg{\langle} \hspace{- 2.3 mm}  \Bigg{\langle} }
\newcommand{\pirBb}{\Bigg{\rangle} \hspace{- 2.3 mm}  \Bigg{\rangle} }
\newcommand{\indexnegspacepirB}{\hspace{-1.2 mm}}
\newcommand{\indexnegspacepirBb}{\hspace{-1.3 mm}}

\newcommand{\A}{A_{\text{eff}}}
\newcommand{\ab}{a}

\newcommand{\Aext}{A}
\newcommand{\af}{a}
\newcommand{\afn}{a_0}
\newcommand{\Aind}{A_{\text{ind}}}
\newcommand{\At}{A_{\text{tot}}}

\newcommand{\Bind}{B _{\text{ind}}}
\newcommand{\Bext}{B_{\text{ext}}}
\newcommand{\Bt}{B _{\text{tot}}}

\newcommand{\calN}{{\cal N}}
\newcommand{\calR}{{\cal R}}
\newcommand{\cd}{{\cal D}}
\newcommand{\Cs}{C_s}
\newcommand{\Ct}{C_t}
\newcommand{\Cinfty}{C_{\infty}}

\newcommand{\D}{\text{D}}

\newcommand{\e}{{e}}
\newcommand{\elop}{\hat{\psi}}

\newcommand{\Ham}{H}
\newcommand{\Hb}{H(-)}
\newcommand{\Hel}{H_{\text{el}}}
\newcommand{\Hf}{H(+)}
\newcommand{\Hint}{H_{\text{int}}}
\newcommand{\Hl}{H^{-}}
\newcommand{\Hn}{H_0}
\newcommand{\Hu}{H^{+}}

\newcommand{\la}{\left \langle} 

\newcommand{\pf}{\frac{d{\bf p}}{(2 \pi \hbar)^2}}
\newcommand{\phit}{\phi _{\text{tot}} }
\newcommand{\psiel}{\psi _{\text{el}}}
\newcommand{\psield}{\bar{\psi} _{\text{el}}}

\newcommand{\ra}{\right \rangle} 
\newcommand{\raa}{\right \rangle_{a , t}} 
\newcommand{\raan}{\right \rangle ^{\text{eq}}_{a , t}} 
\newcommand{\rhoac}{\rho _{a\text{c}}}

\newcommand{\rhoc}{\rho _{\text{c}}}
\newcommand{\rv}{{\bf r}}

\newcommand{\Scom}{S_{\text{ComFer}}}
\newcommand{\SCS}{S_{\text{CS}}}
\newcommand{\Seff}{S_{\text{CF}}}
\newcommand{\Seffa}{S_{\text{eff}}}

\renewcommand{\sp}{s^{\prime}}
\newcommand{\spp}{s^{\prime \prime}}

\newcommand{\T}{k_{\text{B}} T}

\newcommand{\Uu}{U^{+}}
\newcommand{\Ul}{U^{-}}

\newcommand{\vAext}{{\bf A}}
\newcommand{\vA}{{\bf A}_{\text{eff}}}
\newcommand{\va}{{\bf a}}
\newcommand{\vab}{{\bf a}}
\newcommand{\vaf}{{\bf a}}
\newcommand{\vAt}{{\bf A}_{\text{tot}}}
\newcommand{\vda}{\Delta {\bf a}}
\newcommand{\vEind}{{\bf E} _{\text{ind}}}
\newcommand{\vEt}{{\bf E} _{\text{tot}}}
\newcommand{\vdr}{ \Delta {\rv}}
\newcommand{\vj}{{\bf j}}
\newcommand{\vja}{{\bf j}_a}
\newcommand{\vP}{{\bf P}}
\newcommand{\vp}{{\bf p}}
\newcommand{\vq}{{\bf q}}
\newcommand{\vR}{{\bf R}}
\newcommand{\vs}{{\bf r}^{\prime \prime}}
\newcommand{\vv}{{\bf v}}

\newcommand{\wa}{w_a}
\newcommand{\wig}{w}

\begin{document}
\draft
\title{\bf A single-particle path integral for composite
fermions and the renormalization of the effective mass}

\author{Kasper Astrup Eriksen\footnote{Present address: Nordita, 
Blegdamsvej 17, DK-2100 Copenhagen, Denmark}, Per Hedeg{\aa}rd, and Henrik Bruus}
\address{{\O}rsted Laboratory, Niels Bohr Institute APG, Universitetsparken 15, DK-2100 Copenhagen, Denmark}
\date{December 20, 2000}
\maketitle

\begin{abstract}

To study composite fermions around an even denominator fraction,
we adapt the phase space single-particle path integral technique
for interacting electrons in zero magnetic field
developed recently by D.S.\ Golubev and A.D.\ Zaikin, Phys.\ Rev.\ B {\bf 59}, 
9195 (1999).
This
path integral description gives an intuitive picture of composite
fermion propagation very similar to the
Caldeira-Leggett treatment of a particle interacting with an 
external environment. We use the new description to explain the
origin of the famous cancellation between the self-energy and
the vertex corrections in semi-classical  
transport measurements.
The effective range 
of the cancellation is
given by the size of the propagating particle, which for the Coulomb
interaction scales with the temperature $T$ as $T^{-1/4} |\log T|^{-1}$ 
in the semi-classical
limit. Using this scheme we find that the effective mass in the 
semi-classical limit for composite
fermions in GaAs is approximately $6$ times the bare mass.   
\end{abstract}



\section{Introduction}

In this article we address the question of semi-classical
propagation of an interacting system in a real space formulation
\cite{GolubevZaikin}.
In particular we have in mind the composite fermion (CF) system
which is seen in the fractional quantum Hall regime around filling
factor $\frac{1}{2}$. The experimental efforts so far indicates
that this system is to a high degree a classical system in the
sense that it is very hard to observe any kind of interference
effects.

In elementary textbooks \cite{elementarytextbooks} the notion of
semi-classical transport is sometimes introduced through localized
wave packets, and these appear to be essential building blocks.
Historically this has really not always been the case. For
instance the derivation of the Boltzmann equation in the case of
impurity scattering typically only uses individual quasi-particles
with a given momentum i.e.\ delocalized plane waves%
\cite{RammerReview}. This is also the case for other weakly
interacting systems like electron scattering on phonons at low temperatures. 
However, a
Boltzmann equation for electrons scattering on hot phonons could
not be derived using plane waves. Prange and Kadanoff
solved that problem by shifting focus from individual
quasi-particles to deformations of the Fermi surface
\cite{KadanoffPrange}. In real space this step corresponds to
forming an electron wave packet that is localized in the direction
of propagation. This localization in the direction of propagation
causes the electron-phonon scattering to be very brief in time due
to the huge difference between the Fermi velocity and the sound
velocity. Recently Kim, Lee, and Wen showed that the Boltzmann equation 
for composite fermions only
makes sense for smooth Fermi surface deformations \cite{LeeBoltz}.
In real space the wave packet now also needs to be localized in
the direction transverse to the direction of propagation, so the
composite fermions is the first system where the textbook
introduction actually is a proper description. For composite
fermions the necessity of transverse confinement of the wave
packet can be understood intuitively. The important interaction is
mediated through a gauge field and the coupling is thus through
the flux enclosed by an area. For a localized wave packet with a
finite width propagating along a classical path the relevant area
is the area swept by the wave packet. The width of the packet is
thus a natural cut-off on the transverse gauge fluctuations. In this 
article we pursue some of the consequences of this idea.

We will study the real space propagation with the help of phase
space path integrals. Originally, the Feynman path integral
technique was developed for non-interacting systems %
\cite{FeynmanOriginalNonint}. Feynman-Vernon  \cite{FeynmanVernon} and
Caldeira-Leggett \cite{CaldeiraLeggett} expanded the idea to systems 
interacting with an environment. Recently
Golubev and Zaikin extended the technique to cover the
linear response regime for the reduced one-particle density matrix
in a Coulomb interacting gas of electrons \cite{GolubevZaikin}.
This last step is not yet widely recognized. Nonetheless, in Sec.\ %
\ref{density matrix} we adapt it to the case of composite
fermions. We
summarize the main points of the path integral technique
in subsection %
\ref{Subsec Density matrix major results}.
In Sec. \ref{Sec KuboFormulaComparision} the path integral technique is on 
a formal level compared with the Kubo formula description of transport and,
more briefly, with the Landauer-B\"{u}ttiker scattering
approach to transport.
In Sec. \ref{Sec SemiclasSeparation} we discuss the typical semi-classical
approximation scheme of saying that the two paths are almost identical,
and the typical width between the two paths is
calculated. In Sec. \ref{Sec effectivemass} the calculation of the effective
mass of composite fermions is carried through in the semi-classical limit.

We have found it necessary to introduce
quite a few averages and short hand notations. As a readers guide we list 
them here. 
$\la X \ra$ is the full quantum and thermal average. A subscript $0$ 
as in $\la X \ra_0$ means that the
average is performed for the non-interacting electron gas.
$\la X \raa$ is the average over different Chern-Simon gauge field configurations, the precise definition is Eq.\ (\ref{Eq: afield average def}).
$\avell X\averl=\avell X\averl_{[0,t]}$ in Eq.\ (\ref{Eq konjugate def <<>>t})
is the average in the time interval $[0,t]$ over different single-particle paths, where each path is 
given a weight that reflects its importance in a given experiment. It thus 
depends 
on what experiment is considered.
$\pill X \pirl=\pill X \pirl_{[0,t]}$ 
is the shorthand notation for the sum over single-particle paths including the
weight due to the interactions. It is defined in %
Eq.\ (\ref{Eq konjugate def <<<>>>t}).
$\gbl X \gbr$ defined in Eq.\ (\ref{Def gbl gbr shorthand}) is a short hand
notation for the Grassmann integral. It is only used in Appendix %
\ref{AppDiff}.
Quite generally the superscript `$\text{eq}$' refers to the average being
calculated in equilibrium. For instance $\la X \raan$ in Eq.\ (\ref{Eq: afield equil average def}) is the equilibrium version of 
$\la X \raa$.


\section{The density matrix}
\label{density matrix}
We consider a two dimensional electron gas in a static and
homogeneous external magnetic field $\Bext$ perpendicular to the
plane. We write $\Bext$ as a scalar since its direction is fixed,
and likewise we consider the rotation of any vector potential in the
two dimensional plane as a scalar.
Furthermore, we also apply an external electrical potential
$\phi (\rv, t)$. The external potentials are written in four-vector form
as $\Aext = (\phi, \vAext)$.
In this section the goal is to derive a path-integral description of the linear
response of the reduced one-particle density matrix $\rho$ to the potential 
$\phi (\rv, t)$.
For simplicity we will in this section neglect the influence of impurities,
but it is straight forward to include them in the formalism as an extra
external potential.

\subsection{The formulation of the problem using electrons as the fundamental particles}

We will assume the system is described by the Hamiltonian
\begin{equation}
\Hel = \Hn[\Aext] + \Hint .
\end{equation}
Here
\begin{equation}
\Hn[\Aext] = \int d \rv \; \elop ^\dagger (\rv)
\left \{ \frac{1}{2 m}
\left [\frac{\hbar}{i} \nabla + e \vAext (\rv) \right ]^2
- e \phi (\rv, t) -\mu \right \} \elop(\rv)
\label{H0Def}
\end{equation}
is the Hamiltonian for non-interacting particles in the 2D plane
represented by
the second quantized electron field operators
$\elop(\rv)$ and $\elop ^\dagger (\rv)$, while
\begin{equation}
\Hint = \frac{e^2}{2} \int d \rv \int d \rv ^\prime
\elop ^\dagger (\rv) \elop ^\dagger (\rv ^\prime ) v(\rv - \rv ^\prime )
\elop(\rv ^\prime ) \elop(\rv) + H_{\text{eb}}
\label{Eq: Hint Def}
\end{equation}
is the part due to electron-electron interaction. $H_{\text{eb}}$ 
represents the interaction between the
electrons and a fixed neutralizing positively charged background
jellium.
$m$ is the mass of the electron, $\mu $ is the chemical
potential, $-e$ is the electron charge, and $e^2 v(\rv)$ is the
Coulomb pair interaction between electrons. If the screening
effect of the back gate is neglected $v(\rv) = 1/(4 \pi
\epsilon |\rv|)$, where $\epsilon $ is the permittivity of the
medium surrounding the two dimensional electron gas.

All the physical one-particle quantities of the system, like the current
and the density, can be expressed in terms of the reduced one particle density matrix
\begin{equation}
\rhoc(\rv, \rv ^\prime; t ) = \la \elop ^\dagger
(\rv ^\prime, t ) \elop (\rv, t) \ra
\label{Eq: DefDensityMatrixCanonicalEl}
\end{equation}
Here $\la X \ra$ denotes the full quantum mechanical and thermal average
of $X$.
The subscript ${\text{c}}$ stands for canonical,
because upon a Fourier transformation the annihilation and creation operators
correspond to states labeled by the canonical momentum.
This labeling is not gauge invariant so neither is $\rhoc(\rv, \rv ^\prime )$, but
it has some nice formal properties we will make use of in the following.
Final physical expectation values we prefer to phrase in terms of the gauge
invariant density matrix
\begin{equation}
\rho (\rv, \rv ^\prime ) = \rhoc(\rv, \rv ^\prime ) \e ^{ i \frac{e}{\hbar } \int_{\rv ^\prime }^{\rv} d \vs \cdot \vAext(\vs)} ,
\label{Def rho rhoc}
\end{equation}
where the path of integration is chosen to be along the straight line from
$\rv ^\prime $ to $\rv$. In general different choices of the
integration path lead to different values of $\rho$, but we are mainly interested
in local probes like
the density and the current and then any path that locally approaches a straight line
yields the same result. In such local
probes it is convenient to introduce variables for the average
$\vR=\frac{1}{2}(\rv+\rv^\prime)$, and for the difference
$\vdr=\rv-\rv^\prime$. We use a new symbol for the density matrix
expressed in these variables
\begin{equation}
\wig (\vR, \vdr) = \rho \left( \vR + \frac{\vdr}{2}, \vR -
\frac{\vdr}{2} \right).
\label{Def Wigner rho real space}
\end{equation}
The Fourier transform with respect to the difference variable is
the corresponding gauge invariant Wigner distribution function
\begin{equation}
\wig (\vR, \vp) = \int d \vdr \; \e^{- \frac{i}{\hbar} \vp \cdot \vdr}
\wig (\vR, \vdr) .
\label{Def Wigner rho}
\end{equation}
In terms of the gauge invariant Wigner distribution function
the current density at $\vR$ is given by
\begin{equation}
\vj(\vR) = - e \int \pf \frac{\vp}{m} \wig (\vR, \vp).
\end{equation}
Notice the momentum in the Wigner distribution $\vp$ in the
classical local limit is the kinetic momentum.

\begin{figure}
\centerline{\epsfig{file=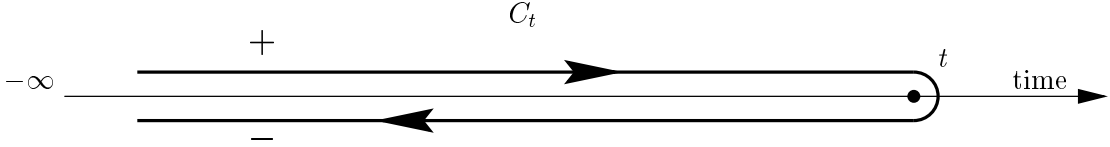, height=40mm}}
\caption{The Grassmann fields 
$ \psield $, $\psiel $, $\bar{\psi } $, and $\psi $ and the 
Chern-Simon gauge field $a = (a_0, \va)$ are all live on the contour $\Ct$. 
As can be seen from the drawing the contour $\Ct$ runs along the physical 
time axis from
$-\infty$ up to time $t$ and then back again to $-\infty$. The contour time
$\sigma$ is on the forward and backward part of the contour denoted 
$\sigma=(s,\fin)$ and $\sigma=(s,bi)$ respectively, where $s$ is the physical
time. To each physical 
time $s<t$ there is thus as an example two independent Chern-Simon gauge fields
$a(s,fin)$ and $a(s,bi)$. If in particular $t=\infty$ the usual Keldysh 
contour $\Cinfty$ is recovered.}
\label{Fig1path}
\end{figure}

The thermal and quantum average of $\rhoc$ and $\rho$ can be calculated
with the help of a path integral over Grassmann fields
$ \psield $ and $\psiel $. $ \psield $ corresponds to the operator
$\elop ^{\dagger}$ and they should not be confused. Likewise, $\psiel $ 
corresponds
to $\elop$. Using this representation we obtain
\begin{equation}
\rho (\rv, \rv ^\prime;t ) = \e ^{ i \frac{e}{\hbar } \int_{\rv
^\prime }^{\rv} d \vs \cdot \vAext(\vs)} \frac{\int \cd \psield
\int \cd \psiel \psield ( \rv ^\prime, t ) \psiel(\rv, t) \exp
\left (\frac{i}{\hbar} S[\psield, \psiel; t]   \right )  } {\int
\cd  \psield \int \cd \psiel \exp\left( \frac{i}{\hbar} S[\psield,
\psiel; t] \right ) },
\label{Eq: GrassRepElecRho}
\end{equation}
where the action $S$ is given by
\begin{equation}
S[\psield, \psiel; t] = \int _{\Ct} d \sigma \int d \rv \; \psield
(\rv, \sigma ) i \hbar \frac{\partial }{\partial \sigma }
\psiel(\rv, \sigma
) - \int _{\Ct} d \sigma \Hel[\psield, \psiel]        .
\label{Action1Grassmann}
\end{equation}
The contour $\Ct$ runs from minus infinity up to the observation
time $t$ and back again to minus infinity.  The contour time
$\sigma=(s,\mu)$, where $s$ is a real time and $\mu= \pm $ is a
contour index that takes the value $\fin$ on the forward part of
the contour and $\bi$ on the backward in time part of the contour, 
see Fig.\ \ref{Fig1path}.
In the prefactor $\psield ( \rv ^\prime, t ) \psiel(\rv, t)$ the
creation Grassmann field $\psield ( \rv ^\prime, t )$ has to come
later than the annihilation Grassmann field $\psiel(\rv, t)$ on
the contour $\Ct$, but whether both fields are on the forward or
backward parts of the contour $\Ct$ or $\psiel(\rv, t)$ is on the
forward part, while $\psield ( \rv ^\prime, t )$ is on the
backward in time part of $\Ct$ is immaterial.

\subsection{The Chern-Simon field dependent density matrix}
\label{udledning}

The first step towards obtaining an effective single-particle path integral 
description is to introduce
fluctuating fields as extra integration variables such that the action becomes quadratic in the
Grassmann fields. The electron physics is then reduced to that of non-interacting electrons in
a fluctuating field background. In the case of zero magnetic field the relevant transformation
is a Hubbard-Stratonovich transformation that introduces an effective electric potential at
each contour point \cite{GolubevZaikin}. In the case of composite fermions the relevant
transformation is a Chern-Simons transformation that introduces a Chern-Simon gauge field
$a = (a_0, \va)$ at each contour point, and that also performs a singular gauge transformation on
the Grassmann fields attaching two magnetic flux quanta
$h/e$ to each electron. The transformed Grassmann fields we denote by $\bar{\psi } $ and $\psi $, not to be confused with $\elop ^{\dagger}$, $\elop$, 
$ \psield $ and $\psiel $.

The action transforms into the standard action for
composite fermions \cite{CFaction}
\begin{equation}
\label{action1CompositeFermion}
\Seff[\bar{ \psi}, \psi,a,\A - a; t] 
=
\SCS[a;t]
+
e \int _{\Ct} d \sigma \int d \rv  a_0(\rv, \sigma ) n + \int
_{\Ct} d \sigma \int d \rv \bar{\psi } (\rv, \sigma ) i \hbar
\frac{\partial }{\partial \sigma } \psi(\rv, \sigma ) \nonumber\\
-
\int _{\Ct} d \sigma \Hn[\A - a].   
\end{equation}
Here 
$\SCS[a;t]$ is the Chern-Simon action
\begin{eqnarray}
\SCS[a ;t] &=& \int _{\Ct} d \sigma \int d \rv \left( \frac{e^2}{4
\pi \hbar} a_0(\sigma ) \nabla \times \va (\sigma ) + \frac{e^2}{8
\pi \hbar} \va(\sigma ) \times \frac{\partial \va}{\partial \sigma
} \right)
\nonumber\\ &&
-\frac{e^4}{2 (4
\pi \hbar )^2} \int _{\Ct} d\sigma \int d \rv \int d \rv ^\prime
\left[\nabla \times \va (\rv, \sigma )\right]
v(\rv - \rv ^\prime )
\left[ \nabla \times \va(\rv ^\prime , \sigma )\right] .
\label{Eq: Def SCS}
\end{eqnarray}
The average mean-field magnetic field $2 n h/ e $ has been
subtracted from both the externally magnetic field $\nabla \times
\vAext$ and the Chern-Simons gauge field. The remaining fields are
denoted by $\A = ( 0 , \vA )  = (0 , \vAext - \left ( \nabla
\times\right )^{-1} 2 \frac{h }{e} n )$ and $a = ( a_0, \va)$
respectively. For notational consistency with the subsection on
linear response we have chosen to have a vanishing external
potential, $\phi=0$. If non-zero it would be added to $\A$. The
last term in $\SCS$ originates from the interaction term $\Hint$,
Eq.\ (\ref{Eq: Hint Def}), after applying the identity $\nabla
\times \va (\rv, \sigma ) = 2 \frac{h}{e} \left ( \psield(\rv,
\sigma ) \psiel(\rv, \sigma ) - n \right )$. For later use we
notice that in the terms that couple the Chern-Simons gauge field
$a$ to the Chern-Simons fields $\bar{ \psi}$ and $\psi$, $a$ only
enters in the combination $\A-a$.

In the first line of Eq.\ (\ref{action1CompositeFermion}) the temporal
integration contour can be extended to the Keldysh contour
$\Cinfty$, since without a Hamiltonian there is no real time development
and consequently the integral over the $a$-field on the Keldysh
contour from the observation time $t$ to infinity is one. $\Cinfty$
is used below for technical reasons whenever the evolution of various
quantities like effective actions and density matrices in a fixed background gauge field
history $a$ is discussed. The contour $\Ct$ is used when the connection
to a physical density matrix as in Eq.\
(\ref{DensityMatrixChernSimonrep1a}) is wanted.

The prefactor in Eq.\ (\ref{Eq: GrassRepElecRho}) transforms as
\begin{equation}
\psield ( \rv ^\prime, t ) \psiel(\rv, t)
\e ^{ i
\frac{e}{\hbar } \int_{\rv ^\prime }^{\rv} d \vs \cdot
\vAext(\vs)} 
\hookrightarrow
\bar{\psi }(\rv ^{\prime} , t) \psi (\rv, t) \e ^{ 
i
\frac{e}{\hbar } \int _{\rv ^{\prime} }^{\rv} d \vs \cdot\left[
\A(\vs, t)-\bar{a}(\vs, t)\right]
}, 
\end{equation}
where 
\begin{equation}
\bar{\va}(\vs, t) =
\frac{1}{2}\left[\vaf(\vs, t, \fin) + \vab(\vs,t, \bi) \right] .
\end{equation} 
is the average of the Chern-Simons gauge field on the two parts of $\Ct$.
This average
value is chosen for notational consistency. It is allowed since the
$a$-field is
continuous on the contour $\Ct$, so it does not matter whether the
Chern-Simons field at the turning point $t$ is evaluated on the
forward or on the backward part of the contour. The
total expression for the density matrix in terms of the
transformed fields is
\begin{equation}
\rho (\rv, \rv ^\prime;t ) = 
\left[\int \cd a \int \cd \bar{\psi }  \int \cd \psi
\e^ { 
\frac{i}{\hbar} \Seff 
}\right]^{-1}
\int \cd a \int \cd \bar{\psi }  \int \cd \psi \;
\bar{\psi } ( \rv ^\prime, t ) \psi(\rv, t)
\e ^{
\frac{i}{\hbar} \Seff + i \frac{e}{\hbar } \int_{\rv ^\prime }^{\rv} d \vs \cdot \left [\vA(\vs) - \bar{ \va} (\vs, t)  \right ] 
}.
\label{DensityMatrixChernSimonrep1a}
\end{equation}
The exponent is quadratic in the Grassmann CF variables $\bar{\psi } $ and $\psi $ and
thus describes non-interacting particles in a gauge field history
$\A - a$. A typical gauge field history is not physically
realizable since the gauge field $a$ at a given time $s$ is
different on the forward and backwards parts of the
contour $\Ct$, $\af(s,\fin) \neq \ab(s,\bi)$. Despite this we will in
analogy with the usual definition of a density matrix Eq.\ (\ref{Eq:
DefDensityMatrixCanonicalEl}) and following Golubev and Zaikin
\cite{GolubevZaikin} define the Chern-Simon field dependent
density matrix
\begin{equation}
\rhoac(\A-a; \rv, \rv ^\prime, t ) =
\left[ \int \cd \bar{\psi }  \int \cd \psi \exp \left (\frac{i}{\hbar}
\Seff   \right ) \right]^{-1}
\int \cd \bar{\psi } \int \cd \psi \;
\bar{\psi } ( \rv ^\prime, t ) \psi(\rv, t) \exp
\left (\frac{i}{\hbar} \Seff \right )
.
\label{Eq: Defrhoac1}
\end{equation}
Notice that the right-hand side only depends on the gauge fields in the
combination $\A-a$, since the contribution from $\SCS[a ;t]$ to
the action $\Seff$ cancels out. The subscript $a$ on $\rhoac$ is
to remind us that $\rhoac$ depends on the field history, while the
subscript $c$ stands for canonical in analogy with $\rhoc$, see
Eq.\ (\ref{Eq: DefDensityMatrixCanonicalEl}). The physical density
matrix is recovered from $\rhoac(\A-a; \rv, \rv ^\prime, t )$ by
averaging over the different $a$-field histories
\begin{equation}
\rho (\rv, \rv ^{\prime} , t) = \la \rhoac(\A-a; \rv, \rv
^{\prime} , t) \exp \left \{ i \frac{e}{\hbar } \int_{\rv ^\prime
}^{\rv} d \vs \cdot \left [\vA(\vs) - \bar{ \va}(\vs, t) \right ]
\right \} \raa. \label{Eq: MainConnRhoRhac}
\end{equation}
In the average
\begin{equation}
\la X \raa = \frac{\int \cd a  \; X
\exp \left (\frac{i}{\hbar} \Seffa[a; t] \right )}
{\int \cd a
\exp \left (\frac{i}{\hbar} \Seffa[a; t]   \right )} ,
\label{Eq: afield average def}
\end{equation}
the weight
\begin{equation}
\exp \left ( \frac{i}{\hbar } \Seffa[a; t] \right ) = \int \cd
\bar{\psi }  \int \cd \psi \exp \left (\frac{i}{\hbar}
\Seff[\bar{\psi } , \psi, a,\A-a; t]   \right ) .
\label{Eq: DefSeffa}
\end{equation}
is the denominator in Eq.\ (\ref{Eq: Defrhoac1}).
The effective action $\Seffa[a; t]$ for the $a$ field also depends on the
external gauge field $\A$, but
this dependence has been suppressed in the notation, since it is
not needed in the following.
Eq.\ (\ref{Eq: MainConnRhoRhac}) is an important formula, since it
establishes the connection between the density matrix $\rhoac$ depending on a given history of
the gauge field
and the physically accessible density matrix $\rho$.

\subsection{The effective action for the $a$-field}

Introducing the $a$-dependent current $\vja(\A-a; \rv, s)$ at time $s$ 
\begin{equation}
\vja(\A-a; \rv, s) = \left. \frac{- e}{2 m}\left [ \frac{\hbar
}{i} \frac{\partial }{\partial \rv} + e \vA (\rv) - e \bar{\va}
(\rv)- \frac{\hbar }{i} \frac{\partial }{\partial \rv ^\prime } +
e \vA (\rv ^\prime ) - e \bar{\va} (\rv ^\prime ) \right ] \rhoac
(\rv, \rv ^\prime; s ) \right|_{\rv ^\prime = \rv},
\label{Eq: CurrentPresenceGaugeFieldHistory}
\end{equation} 
it is shown in appendix \ref{AppDiff} 
that the effective single-particle action $\Seffa[a;t] $ is given by
\begin{equation}
\Seffa[a;t] = \SCS[a;t] - \int _{-\infty}^{t} d s \int d \rv \left
[ e \Delta a_0(\rv, s) (\rhoac(\rv, \rv; s ) - n) + \vda(\rv, s)
\cdot \vja(\rv, s) \right ] , \label{Eq: EqSeffa Generel}
\end{equation}
where
\begin{equation}
\Delta a_{\alpha} (\rv, s)= a_{\alpha}(\rv,s,\fin )- a_{\alpha}(\rv, s,\bi )
\end{equation}
is the difference of the Chern-Simons gauge field on the two parts of $\Ct$.
$\SCS[a;t]$ is the usual Chern-Simons action, see Eq.\ (\ref{Eq: Def SCS}).
In the notation the explicit dependence
of the density matrix $\rhoac$ and the current $\vja$ 
on the gauge field $\A-a$ has
been suppressed.
In equilibrium, which is denoted by the superscript
"${\text{eq}}$", we have to second order in the $a$-field
\begin{eqnarray}
\Seffa ^{\text{eq}}[a;t] &\approx& \SCS[a;t] - \sum_{\alpha
\,,\,\beta}\int _{-\infty}^{t} d \sp \int _{-\infty}^{\sp} d \spp
\int d \rv ^{\prime} \int d \rv ^{\prime \prime}
\Delta a_\alpha(\rv^{\prime}, \sp)  R_{\alpha \beta}(\rv^{\prime}, \sp ;
\rv^{\prime \prime}, \spp) \bar{a} _{\beta}(\rv^{\prime\prime},
\spp) \nonumber\\
&&
+ \frac{i}{2}\sum_{\alpha
\,,\,\beta} \int _{-\infty}^{t} d \sp \int
_{-\infty}^{t} d \spp \int d \rv^{\prime} \int d \rv^{\prime}
\Delta a_\alpha(\rv^{\prime}, \sp)  F_{\alpha \beta}(\rv^{\prime}, \sp ;
\rv^{\prime \prime}, \spp) \Delta a _{\beta}(\rv^{\prime\prime},
\spp) .
\end{eqnarray}
The summation variables $\alpha,\beta \in \{0,x,y\}$ run over the single
time-like and the two spatial indices e.g. $a_{\alpha}=(a_0,a_x,a_y)$.
The terms linear in $a$ vanishes, since $\rhoac(\A, \rv, \rv, s) =
n$ and $ \vja(\A; \rv, s)= \bnul$. The retarded $3\times 3$
response matrix
$\left(\R\right)_{\alpha \beta}$ is given by the usual Kubo
formula
apart from an 
extra minus in the part that mixes the density with the currents.
\begin{eqnarray}
\R(\rv^{\prime}, \sp ; \rv^{\prime \prime}, \spp) &\equiv&
-\frac{i}{\hbar }
\left(\matrix{ \la \left[
-e \rhoac(\rv^{\prime}, \sp ) ,
-e \rhoac(\rv^{\prime \prime}, \spp )\right] \ra _0
&
-\la\left[-e \rhoac(\rv^{\prime}, \sp ) ,
\vja(\rv^{\prime\prime}, \spp ) \right] \ra _0
\cr
-\la\left[\vja(\rv^{\prime}, \sp) ,
-e \rhoac(\rv^{\prime \prime}, \spp )\right] \ra _0
&
\la\left[\vja(\rv^{\prime}, \sp) ,
\vja(\rv^{\prime\prime}, \spp ) \right] \ra _0}
\right)
\Theta ( \sp - \spp)
\nonumber\\ && 
+\left(\matrix{0 & 0 \cr 0 & \frac{e^2}{m}\rhoac(\rv^{\prime}, \sp)
\delta(\rv^{\prime}-\rv^{\prime \prime})\delta(\sp - \spp)}\right)
\label{Def R matrix}
\end{eqnarray}
and the so-called fluctuation kernel $\left(\F\right)_{\alpha \beta}$ by
\begin{equation}
\F(\rv_1, s_1 ; \rv_2, s_2)
\equiv
\frac{1}{2 \hbar } \left(\matrix{ \la \left\{
-e \rhoac(\rv^{\prime}, \sp ) ,
-e \rhoac(\rv^{\prime \prime}, \spp )\right\} \ra _0
&
-\la\left\{-e \rhoac(\rv^{\prime}, \sp ) ,
\vja(\rv^{\prime\prime}, \spp ) \right\} \ra _0
\cr
-\la\left\{\vja(\rv^{\prime}, \sp) ,
-e \rhoac(\rv^{\prime \prime}, \spp )\right\} \ra _0
&
\la\left\{\vja(\rv^{\prime}, \sp) ,
\vja(\rv^{\prime\prime}, \spp ) \right\} \ra _0}
\right) .
\label{Def F matrix}
\end{equation}
The dependence of the density matrix $\rhoac$ and the current
$\vja$ on the gauge field $\A-a$ has been suppressed in the notation.
$\la X\ra_0$ means the average over the non interacting electron gas without
the Chern-Simons gauge field $a$, and $\Theta $ is the step function.
In equilibrium the retarded response matrix $\R$ is related to the
fluctuation matrix $\F$ through the fluctuation dissipation theorem
\begin{equation}
F _{\alpha \beta } ( \rv_1 , \rv_2; \omega)  =
\frac{i}{2} \coth \left ( \frac{\hbar \omega }{2 \T} \right )
\left [ R _{\alpha \beta } ( \rv_1 , \rv_2; \omega) -
R _{\beta \alpha } ( \rv_2 , \rv_1; -\omega) \right ] ,
\label{Fluctuation Dissipation Ori}
\end{equation}
where $\omega $ is the Fourier frequency corresponding to
the time difference $\sp - \spp$. It is used that in equilibrium
$\R$ and $\F$
only depends on $\sp -
\spp$. In terms of the $2\times2$ conductivity tensor
$\bar{\bar{\sigma}}(\vq,
\omega )$ the response matrix $\R$ is  written as
\begin{equation}
\R(\vq, \omega ) = i \left ( \matrix{ \frac{1}{\omega }
\vq \cdot \bar{\bar{\sigma}}(\vq, \omega) \cdot \vq
& - \vq \cdot \bar{\bar{\sigma}}(\vq, \omega) \cr
- \bar{\bar{\sigma}}(\vq, \omega) \cdot \vq
 & \omega \bar{\bar{\sigma}}(\vq, \omega)}
\right ).
\end{equation}
Below we are mainly going to need the retarded gauge field
propagator $D_{\alpha\beta}$ defined
by%
\begin{equation}
D_{\alpha\beta}\left( \rv^{\prime} ,s^{\prime};\rv^{\prime\prime}
,s^{\prime\prime}\right) 
\equiv 
-\frac{i}{\hbar}e^{2}\la
\bar{a}_{\alpha }\left(  \rv^{\prime} ,s^{\prime}\right) \Delta
a_{\beta}\left( \rv^{\prime\prime}  ,s^{\prime\prime}\right) \raan
\label{Def D matrix}%
=
-\frac{i}{\hbar} \Theta\left(s^{\prime}-s^{\prime \prime}
\right) e^2\la \left[a_{\alpha }\left( \rv^{\prime}
,s^{\prime}\right), a_{\beta}\left( \rv^{\prime\prime}
,s^{\prime\prime}\right) \right] \ra_0
\end{equation}
and the correlation function $C_{\alpha\beta}$ for fluctuations of the field
\begin{equation}
C_{\alpha\beta}\left( \rv^{\prime} ,s^{\prime};\rv^{\prime\prime}
,s^{\prime\prime}\right) 
\equiv 
\frac{e^{2}}{\hbar}\la
\bar{a}_{\alpha }\left( \rv^{\prime} ,s^{\prime}\right)
\bar{a}_{\beta}\left( \rv^{\prime\prime}  ,s^{\prime\prime}\right)
\raan
\label{Def C matrix}%
=
\frac{e^{2}}{\hbar}\frac{1}{2}\la \left\{ a _{\alpha }\left(
\rv^{\prime}  ,s^{\prime}\right) , a_{\beta}\left(
\rv^{\prime\prime}  ,s^{\prime\prime}\right)  \right \} \ra_0 %
,
\end{equation}
while the auto-correlation function for $\Delta a_\alpha$ vanishes:%
\begin{equation}
\la \Delta a_{\alpha}\left(  \rv^{\prime}  ,s^{\prime}\right)
\Delta a_{\beta}\left( \rv^{\prime\prime}
,s^{\prime\prime}\right) \raan =0 .
\end{equation}
The average
\begin{equation}
\la X \raan = \frac{\int \cd a  \; X \exp \left (\frac{i}{\hbar}
\Seffa ^{\text{eq}}[a; t] \right )} {\int \cd a \exp \left
(\frac{i}{\hbar} \Seffa ^{\text{eq}}[a; t]   \right )} ,
\label{Eq: afield equil average def}
\end{equation}
is the average $\la X\raa$ of Eq.\ (\ref{Eq: afield average def}),
calculated in equilibrium.

In the Coulomb gauge the most troublesome contributions arises
from the low frequency limit of the so-called transverse part of
the gauge field propagator $D_{\perp\perp}$ \cite{SingularTransverse,HLR}, which in Fourier
space is given by%
\begin{equation}
D_{\perp\perp}\left(  \vq ,\omega\right)  =\sum_{ijkl}\varepsilon
^{ik}\varepsilon^{jl}\frac{q_{i}q_{j}}{q^{2}}D_{kl}\left(  \vq %
,\omega\right) .
\end{equation}
Here $\varepsilon^{xx}=\varepsilon^{yy}=0$,
$\varepsilon^{xy}=-\varepsilon^{yx}=1$, and the indices $i$, $j$,
$k$ and $l$ run over `x' and `y'. %
In equilibrium in the limit $\hbar\omega\ll q
v_{\mathrm{F}}$ the propagator $D_{\perp\perp}$ is given by the standard approximation
\cite{HLR}%
\begin{equation}
D_{\perp\perp}\left(  \vq ,\omega\right)    \approx 
2\pi\hbar\frac {q}{p_{\mathrm{F}}}\frac{1}{i\omega-\frac{2\pi
e^{2}v\left( q\right) q^{3}}{\left(  4\pi\hbar\right)  ^{2}\hbar
p_{\mathrm{F}}}}
= 
2\pi\hbar\frac{q}{p_{\mathrm{F}}}\frac{1}{i\omega-\g_{\eta}%
q^{\eta+1}}  .%
\label{Eq: GaugePropTransversePropInfrared}
\end{equation}
For the Coulomb interaction $v\left(  q\right)
=\frac{\hbar}{2\varepsilon q}$
so $\eta=1$ and $\gc= %
\frac{1}{\hbar p_{\mathrm{F}}^{2}}\frac{1}{2}\sqrt{\pi}\coulener$ in terms of 
the average Coulomb energy $\coulener=\frac{e^2}{4 \pi
\permittivity}\sqrt{n}
$. $n$ is the density of the electron gas.
Other values of $1\leq\eta<2$ corresponds to a power law decay
$r^{-\eta}.$

The fluctuation correlation function $\bC$ is given by%
\begin{equation}
\bC\left(  1,2\right)  = \int d3\int d4\,\bD\left( 1,3\right)
\bF\left(  3,4\right)  \bD\left( 2,4\right)  ,
\end{equation}
where $1$,$2$,$3$ and $4$ each are a short hand notation for a
full set of variables $\alpha$,$\rv$,$s$. In equilibrium $\bC$ is
determined
from the fluctuation dissipation theorem %
(\ref{Fluctuation Dissipation Ori})%
\begin{equation}
C_{\alpha\beta}\left(  \rv^{\prime} ,\rv^{\prime\prime};\omega
\right)    =\frac{i}{2}\coth\left(  \frac{\hbar\omega}{2k_{\mathrm{B}}%
T}\right)  \left[  D_{\alpha\beta}\left(  \rv^{\prime}  %
,\rv^{\prime\prime}  ;\omega\right)  -D_{\beta\alpha}\left(
\rv^{\prime\prime}  ,\rv^{\prime}  ;-\omega\right)
\right] %
\label{Eq Fluc Diss Theorem General 3}
\end{equation}
In particular in the Coulomb gauge is the most singular
contribution due to the transverse part of the propagator
\begin{equation}
C_{\alpha\beta}\left(  \vq ,\omega\right) \approx -\left(
\delta_{\alpha\beta}-\frac{q_{\alpha}q_{\beta}}{q^{2}}\right)
\coth\left( \frac{\hbar\omega}{2k_{\mathrm{B}}T}\right) \Im
D_{\perp\perp }\left(  \vq ,\omega\right)
\end{equation}

\subsection{The equation of motion for the gauge field dependent density matrix}

In appendix \ref{AppDiff}
we have by brute force differentiation showed that $\rhoac$ fulfills the
all important nonlinear equation of motion
\begin{equation}
i \hbar \pderivet \rhoac = (1 - \rhoac )
\Ham(t,\fin) \rhoac - \rhoac \Ham(t,\bi) (1 - \rhoac ) . %
\label{Diff rhoac}
\end{equation}
This equation should be read as an operator identity, i.e.\ we have
suppressed all arguments and the spatial integrals over the
internal variables, $1$ is a shorthand notation for the
identity operator $\delta(\rv^{\prime}-\rv^{\prime \prime \prime})$,
$\Hf$ is the kernel in the quadratic action for the
Grassmann CF variables on the forward part of the contour, i.e.
\begin{equation}
\Ham (\rv^{\prime}, \rv^{\prime \prime }; t,\fin) = \delta (
\rv^{\prime} - \rv^{\prime \prime })  \left \{ -
e \phi (\rv^{\prime \prime }, t) + e \afn(\rv^{\prime \prime },
t,\fin) + \frac{1}{2 m}\left [\frac{\hbar }{i} \nabla^{\prime \prime
} + e \vA(\rv^{\prime \prime }) - e \vaf(\rv^{\prime \prime },
t,\fin) \right ]^2 \right \} .
\end{equation}
$\af(\fin)=[\afn(\fin),\vaf(\fin)]$ is the gauge field on the forward contour.
Likewise is $\Hb$
the kernel on the backward in time part of the contour. If the
gauge field is identical on the forward and backward sections of
the contour Eq.\ (\ref{Diff rhoac}) specializes to the usual
equation of motion for the density matrix 
$i \hbar \pderivet \rho = [H,\rho]$. The extra complication in the
general case is the term quadratic in the density matrix: $- \rhoac
[ \Hf - \Hb] \rhoac $. However we restrict ourselves to the linear
response regime.

\subsection{Formulation of linear response}

We are interested in the response of the system to an external
driving field $\phi $ and in particular in the change in the
density matrix: $\delta \rho (\rv, \rv ^\prime;t ) = \rho (\rv,
\rv ^\prime;t ) - \rho ^{\text{eq}} (\rv, \rv ^\prime;t )$. The
superscript "${\text{eq}}$" still denotes equilibrium where $\phi
= 0$. In general both the gauge field dependent density matrix
$\rhoac(\A-a; \rv, \rv ^{\prime} , t)$ and the effective action
for the Chern-Simons field $\Seffa[a; t]$ depends on $\phi $,
giving rise to two different contributions to $\delta \rho (\rv,
\rv ^{\prime} , t)$ --- compare with Eqs.\ (\ref{Eq: MainConnRhoRhac})
and (\ref{Eq: afield average def}). The contribution coming from the variation of
$\Seffa[a;t]$ can be accounted for by a change of variable from the
externally applied field $\phi $ to the total gauge field felt be
the composite fermions
\begin{equation}
\At(a;t) = (\phi(t), \bbox{0}) + \Aind(a;t).
\end{equation}
Here $\Aind$ is the self-consistently induced gauge field due to the change
in the current and the density when $\phi $ is applied.
Technically it is defined by the demand that
\begin{equation}
\Seffa[a- \Aind(a);t] = \Seffa^{\text{eq}}[a;t] .
\end{equation}
From this definition it is not immediately clear that the induced field is the same
on the forward and backward part of the contour and thus can be interpreted as a
physical field. However a careful inspection of
Eqs.\ (\ref{Eq: Def SCS}) and
(\ref{Eq: EqSeffa Generel}) shows this is the case.
Furthermore it is seen that the induced electromagnetic
fields corresponding to $\Aind$ are
\begin{eqnarray}
\label{Eq InducedBfield}
\Bind(\A-a; \rv, t) &=& -\frac{4 \pi \hbar }{e} \delta \rhoac(\A-a;
\rv, \rv ; t) \\
\vEind(\A-a; \rv, t) &=& \frac{4 \pi \hbar
}{e^2} \hat{z} \times \delta \vja(\A-a; \rv, t) + e \nabla \int d \rv
^\prime \; v(\rv - \rv ^\prime ) \delta \rhoac( \A-a; \rv,
\rv; t) ,
\label{Eq InducedEfield}
\end{eqnarray}
where
\begin{eqnarray}
\delta \rhoac(\A-a; \rv, \rv ^{\prime} , t) &=& \rhoac(\A +
\At - a; \rv, \rv ^{\prime} , t) \e ^{i\frac{e}{\hbar}
\int_{\rv ^{\prime}}^{\rv}d\rv ^{\prime \prime}\cdot \vAt}
- \rhoac^{\text{eq}}(\A-a; \rv,
\rv ^{\prime} , t) \\
\delta \vja(\A-a; \rv, t) &=& \vja(\A +
\At -a; \rv, t) - \vja^{\text{eq}}(\A-a ; \rv, t)
\end{eqnarray}
are the changes in the density and the current respectively, induced by
the total field $\At$. The phase factor $\exp\left[i\frac{e}{\hbar}
\int_{\rv ^{\prime}}^{\rv}d\rv ^{\prime \prime}\cdot \vAt\right]$ has
been introduced in order to let $\delta \rhoac$ refer to the equilibrium
state. For instance $\delta \vja$ results, when $\delta \rhoac$ is
substituted for $\rhoac^{\text{eq}}$ in the expression for the
equilibrium current Eq.\ (\ref{Eq: CurrentPresenceGaugeFieldHistory}).
Eqs.\ (\ref{Eq InducedBfield}) and
(\ref{Eq InducedEfield}) are what is expected on physical
grounds. There is a magnetic field associated with a change in
the density of composite fermions and thus the amount of attached
flux. There is also a Faraday electric field associated with a current
of attached flux. The last term in Eq.\ (\ref{Eq InducedEfield}) is due
to the electron-electron interaction.

The Jacobian associated with the
variable change from $a$ to $a - \Aind(a)$ in Eqs.\ (\ref{Eq: MainConnRhoRhac})
and (\ref{Eq: afield average def}) is the identity $1$, since both
the density and the current, and thus $\Aind(a)$, are retarded functions of
the gauge field $\A-a$. Consequently the change in
the physical density matrix
\begin{equation}
\delta \rho (\rv, \rv ^\prime;t ) =
\la \delta \rhoac (\A-a,\rv,
\rv ^\prime;t ) \exp \left \{i \frac{e}{\hbar } \int_{\rv ^\prime
}^{\rv} d \vs \cdot \left [\vA(\vs) - \bar{ \va}(\vs, t) \right ]
\right \} \raan
\label{ChangeDensityMatrixConnection}
\end{equation}
is the average of the change in the $a$-dependent density matrix
over the different
$a$-field histories with the weight $\exp\left[
\frac{i}{\hbar } \Seffa^{\text{eq}}[a] \right]$ pertaining to equilibrium.

Golubev and Zaikin \cite{GolubevZaikin} did not perform this
change of variables, since the difference between the external and
the total field is not of great bearing in their application of the
formalism to weak localization. The second term in their formula
(46) which they neglect is exactly what is covered by this change
of variable. For composite fermions such a neglect is not permissible,
since the extra term originating from the variation of $\Seffa[a;t]$
gives rise to for instance the Hall conductance.
In our formulation the Hall effect arises from the electric field induced by
the current --- the first term in Eq.\ (\ref{Eq InducedEfield}).
The mechanism is completely analogous to the treatment in the case of the
Boltzmann equation \cite{Simon in CF book}.

Notice that so far no linearization in the applied field has been
performed and everything is exact.

The linearized equation of motion for $\delta \rhoac(\A-a; \rv, \rv ^\prime,
t) $ is obtained by linearizing Eq.\ (\ref{Diff
rhoac}) in the total external field
\begin{equation}
i \hbar \pderivet \delta \rhoac(\rv , \rv
^\prime ; t) \simeq 
\int d \rv ^{\prime \prime} \Hu(\rv, \rv ^{\prime
\prime}, t)  \delta \rhoac(\rv ^{\prime \prime}, \rv  ^\prime ;t)
- \int d \rv ^{\prime \prime} \delta \rhoac(\rv , \rv ^{\prime
\prime};t) \Hl(\rv ^{\prime \prime}, \rv ^\prime ,t) 
+ i \hbar \D(\rv, \rv ^\prime ,t) ,
\label{LinearDensityMatrixDiff}
\end{equation}
where the explicit dependence on the gauge fields $\A-a$ and $\At$ is
suppressed in
the notation. New one-particle Hamiltonians
\begin{equation}
\Hu = \left .\frac{\Hf + \Hb}{2}\right |_{\phi = 0} +
\left . \left(\frac{1}{2} -  \rhoac \right) %
\left [\Hf - \Hb \right ] \right |_{\phi = 0}
\end{equation}
\begin{equation}
\Hl = \left .\frac{\Hf + \Hb}{2}\right |_{\phi = 0} - \left. \left
[\Hf - \Hb \right ]
 \left(\frac{1}{2} -  \rhoac \right) \right |_{\phi = 0}
\end{equation}
have been introduced in analogy with the equation of motion for the
physical one-particle density matrix
$i \hbar \pderivet \rho=[H,\rho]$. $\Hu \neq \Hl$ because the
Chern-Simons field is different on the forward and backward
contours. The inhomogeneous stimulus $\D = \D_1 + \D_2$ is
split into two parts, each of which are gauge invariant in
the total applied field $\At$. $\D_2$ contains all the dependence
upon the difference gauge field $\Delta \va$, while $\D_1$ is
independent of $\Delta \va$. In Appendix %
\ref{Sec: appDriveD listed} the explicit expressions for $\D_1$
and $\D_2$ are listed.
In linear response $\D_1$ and $\D_2$
contribute additively
to the physical density matrix.
The response due to
$\D_2$ we will neglect, since it vanishes in the
approximation where
$\D$ is replaced by its average value $\la \D
\raan$. Below we do not go beyond this approximation.

In the real space representation used above $\D$ is not gauge
invariant in $\A(\rv) - \bar{\va}$, so we find it convenient
to change representation
to the gauge invariant Wigner function set of variables. In
analogy with Eqs.\ (\ref{Eq: DefDensityMatrixCanonicalEl}), (\ref{Def rho
rhoc}) and (\ref{Def Wigner rho})
\begin{equation}
\D(\rv, \rv ^\prime ,t)
= \int \pf \e^{ 
\frac{i}{\hbar } \int _{\rv ^\prime }^{\rv} d \vs
\cdot \left [ \vp - e \vA(\vs) + e \bar{ \va}(\vs, t) \right ] 
}
\D\left[\frac{1 }{2}(\rv + \rv ^\prime), \vp \right]
.
\end{equation}
When the generalized density matrix $\rhoac(\rv,\rv^{\prime})$ is
peaked around $\rv=\rv^{\prime}$ on a length scale which is much
smaller than the typical scale of variation of the total driving
field, the gauge invariant $\D_1$ and thus $\D$ is
approximately equal to the
standard Boltzmann equation driving term
\begin{equation}
\D (\vR, \vp ) \approx - e\left( \frac{1}{m}\vp \times \unitz \Bt +  \vEt \right)
\cdot \frac{\partial }{\partial \vp}\wa(\A-a; \vR, \vp)  .
\label{DrivingTermapproximation}
\end{equation}
In the presence of the gauge field $\A - a$
the Wigner distribution function $\wa (\A- a; \vR, \vp, t)$ is 
defined in analogy
with Eqs.\ (\ref{Def rho rhoc}) and (\ref{Def Wigner rho})
\begin{equation}
\wa (\A-a; \vR, \vp, t) 
=
\int d \vdr \e ^{ 
-\frac{i}{\hbar} \int _{\vR - \frac{\vdr}{2} }^{\vR +
\frac{\vdr}{2} } d \vs \cdot \left [ \vp - e \vA(\vs) + e \bar{
\va}(\vs, t) \right ] 
}
\rhoac (\A-a;\vR + \frac{1}{2}\vdr, \vR - \frac{1}{2}\vdr; t) .
\end{equation}
In the semi-classical limit it corresponds to a field dependent phase space
distribution function. We will not consider effects beyond the approximation
where it is the Fermi distribution function.

\subsection{The single-particle path integral in linear response}

In this subsection we obtain in terms of a single-particle path
integral the formal solution of $\delta \rho$ in the linear
response i.e.\
Eq.\ (\ref{LinearDensityMatrixDiff}).
The first step is to
notice that the solution to Eq.\ (\ref{LinearDensityMatrixDiff}) can be written
in terms of evolution operators $U^{\mu}$
\begin{equation}
\delta \rhoac (\rv, \rv ^\prime ; t) = - \frac{i}{\hbar }
\int_{- \infty}^{t} d t _i
\int d \rv _1 \int d \rv _2 \;
\Uu(\rv, t, \rv_1, t _i ) \D(\rv_1, \rv _2, t_i)  
\\Ul(\rv_2, t _ , \rv ^\prime, t).
\end{equation}
The evolution operators
\begin{equation}
U^{\mu}(t, t ^\prime ) = \hat{T}^{\mu} \exp \left [ -
\frac{i}{\hbar } \int_{t ^\prime }^{t} d t ^{\prime \prime}
H^{\mu} (t ^{\prime \prime}) \right ], \quad \mu=\pm ,
\end{equation}
where $\hat{T}^+$ and $\hat{T}^{-}$ are the time
and anti-time ordering operators respectively.

The corresponding change in the physical Wigner function is
\begin{equation}
\delta w ( \vR, \vp , t ) = \int_{- \infty}^{t} d t_i \int d\vR _i \int \frac{d \vp _i}{(2 \pi \hbar)^2}
\la J(\vR, \vp ,t; \vR _i , \vp _i, t _i ) \D(\vR _i , \vp _i, t _i )
\raan , %
\label{Eq. Full prop J}
\end{equation}
where
\begin{eqnarray}
J(\vR, \vp ,t; \vR _i , \vp _i, t _i ) 
&=&
\int d \vdr \int d \vdr
_i \int d \rv _i  ^\prime \e ^{ 
- \frac{i}{\hbar } \int
_{\vR + \frac{\vdr}{2}} ^{ \vR - \frac{\vdr}{2} } d \vs \cdot \left [
\vp - e \A( \vs) + e \bar{ \va} (\vs, t) \right ] 
}
\e ^ { 
\frac{i}{\hbar } \int _{\vR _i + \frac{\vdr _i}{2}}
^{ \vR _i - \frac{\vdr _i}{2} } d \vs \cdot \left [ \vp _i - e \A( \vs)
+ e \bar{ \va} (\vs, t _i) \right ] 
} \nonumber\\
&&
\times
J(\vR +
\frac{\vdr}{2},\vR - \frac{\vdr}{2},t;\vR + \frac{\vdr
_i}{2},\vR_i - \frac{\vdr _i}{2},t_i)
\end{eqnarray}
is the propagator of the stimulus $\D$ in the Wigner
representation. In the real space representation
\begin{equation}
J(\vR + \frac{\vdr}{2},\vR - \frac{\vdr}{2},t;\vR + \frac{\vdr
_i}{2},\vR_i - \frac{\vdr _i}{2},t_i) 
=
\Uu(\vR +
\frac{\vdr}{2}, t ; \vR_i + \frac{\vdr _i}{2}, t _i ) \Ul(\vR _i -
\frac{\vdr _i}{2} , t _i; \vR - \frac{\vdr}{2} , t) .
\end{equation}

The evolution operators $\Uu $ and $\Ul$ can be represented as usual 
single-particle phase space path integrals
\begin{equation}
U ^{\mu}(\rv _1, t _1 ; \rv _2 , t  _2  )
=
\int _{\rv (t _2 ) = \rv _2 } ^ {\rv (t _1 ) = \rv _1 } \cd \rv
\int \cd \vp  \exp \left \{ \frac{i}{\hbar } \int _{t _2} ^{t_1} d
t \left[(\vp - e \A + e \bar{ \va} ) \cdot \dot{\rv} -
H ^{\mu} ( \rv(s), \vp(s) , s) \right ] \right\} ,
\label{Eq: PathU}
\end{equation}
where $\mu = \pm $. The Hamiltonian kernels are given by
\begin{eqnarray}
\Hu(\rv, \vp, s)
&=& \frac{\vp^2 }{2 m} + \frac{e ^2 \vda^2(\rv, s) }{8 m} - e \phi
(\rv) + e \bar{a}_0(\rv, s) \nonumber\\
&&
+e \left[\frac{1}{2} - \wa( \rv, \vp(s^+), s ) \right] \left (\Delta a _0 -
\frac{\vp}{m} \cdot \vda(\rv, s) \right ) + i \hbar \frac{e
\vda}{2 m} \cdot \nabla \wa(a; \rv, \vp(s^+), s )
\\
\Hl(\rv, \vp, s)
&=& \frac{\vp^2 }{2 m} + \frac{e ^2 \vda^2(\rv, s) }{8 m} - e \phi
(\rv) + e \bar{a}_0(\rv, s) \nonumber\\&& 
-e \left [ \Delta a _0
- \frac{\vp}{m} \cdot \vda(\rv, s) \right ]
\left[\frac{1}{2} - \wa( \rv, \vp(s^+), s ) \right]
- i \hbar \frac{e \vda}{2 m} \cdot \nabla
\wa(\rv, \vp(s^+), s ) .
\end{eqnarray}
Here $s^+$ on both the forward and backward in time contour is a
bit closer to the observation time $t$ than $s$. The only place
where this small time shift is important is in the Wigner
distribution function $\wa(\rv, \vp(s^+), s)$, which is likely to be
a rapidly varying function of the momentum $\vp$ at the Fermi
surface. Note also that $\vp$ in the
semi-classical limit is the kinetic momentum in the gauge field
$\A - \bar{a} $.

\begin{figure}
\centerline{\epsfig{file=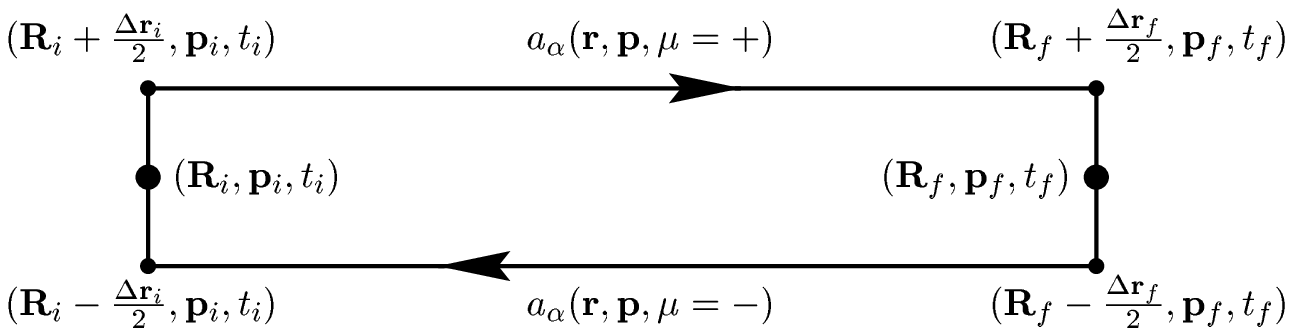, height=25mm}}
\caption{The closed contour path $(\rv(\sigma ), \vp( \sigma ) )$ starts
out at $\sigma = (t _i,\fin)$ at $(\vR_i + \frac{\vdr_i}{2}, \vp _i
)$ and runs along $(\rv(s,\fin), \vp (s,\fin))$ up to $(\vR +
\frac{\vdr}{2}, \vp  )$ at contour time $\sigma =(t,\fin)$. Here it runs along
the straight line path from $\vR + \frac{\vdr}{2}$ to $\vR -
\frac{\vdr}{2}$ with the constant momentum $\vp$ and at constant
$\sigma $. From here it then runs back along $[\rv(s,\bi), \vp
(s,\bi)]$ to $(\vR_i - \frac{\vdr_i}{2}, \vp_i  )$ at contour time $\sigma
=(t_i,\bi)$. Finally it runs along a straight line path back to $\vR_i
+ \frac{\vdr_i}{2}$ with the constant momentum $\vp_i$ and at a
constant time $t_i$}
\label{Fig: Closed response path}
\end{figure}

Using the path integral representation Eq.\ (\ref{Eq: PathU}) for $\Uu$ and $\Ul$ we finally arrive at the following contour path integral representation for the propagator $J$
\begin{equation}
J(\vR, \vp ,t; \vR _i , \vp _i, t _i ) 
=
\int d \vdr \int d \vdr _i \int d \rv _i  ^\prime
\int \cd \rv (\fin) \int \cd \vp (\fin) \int \cd \rv (\bi) \int \cd \vp (\bi)
\exp \left ( \frac{i}{\hbar } S_0 + \frac{i}{\hbar } \Delta S \right ) ,
\label{Eq: PropJumidlet2}
\end{equation}
where
\begin{equation}
S _0 = \oint d \sigma \dot{\rv} \cdot \left \{\vp(\sigma ) - e
\A[\rv(\sigma), s] \right \} - \int _{t _i} ^{t} ds \Hn[\rv (s, \fin),
\vp (s, \fin), s] + \int _{t _i} ^{t} ds \Hn[\rv(s, \bi), \vp(s, \bi), s]
\label{Golchern def of S0}
\end{equation}
is the action for non-interacting particles in an external gauge field $\A$. 
The closed contour is depicted in Fig. \ref{Fig: Closed response path}. 
%
%
All the contributions related to the Chern-Simon gauge field are gathered in the second term in the
exponent of Eq.\ (\ref{Eq: PropJumidlet2})
\begin{eqnarray}
\Delta S &=& + e \oint \left \{ d \rv \cdot \bar{a}[ \rv(\sigma ),
s ] - d \sigma \bar{a} _0 [ \rv(\sigma ), s ] \right \} 
\nonumber\\ 
&& +
e \int _{t _i} ^{t} ds
\left\{ \frac{1}{2} - \wa[ \rv(s,\fin), \vp(s^+,\fin), s ] \right\}
\left \{ \frac{\vp (s, \fin) }{m} \cdot
\vda [\rv(s,\fin), s] - \Delta a_0 [\rv(s,\fin), s]\right \} 
\nonumber\\ %
&& %
+
e \int _{t _i} ^{t} ds
\left\{\frac{1}{2} - \wa[ \rv(s,\bi), \vp(s^+,\bi), s ] \right\} 
\left \{ \frac{\vp (s,\bi) }{m} \cdot
\vda  [\rv(s,\bi), s] - \Delta a_0 [\rv(s,\bi), s]\right \}
\nonumber\\
&&
-
\frac{e^2}{8 m} \int _{t _i} ^{t} ds \left \{\vda[\rv(s,\fin), s]^2 - 
\vda[\rv(s,\bi), s]^2 \right \}
\label{Eq: DS1}
\end{eqnarray}

Below we only consider the approximation obtained by
factorizing Eq.\ (\ref{Eq. Full prop J})
\begin{equation}
\delta w ( \vR, \vp , t ) = \int_{- \infty}^{t} d t_i \int d\vR _i \int \frac{d \vp _i}{(2 \pi \hbar)^2}
\la J(\vR, \vp ,t; \vR _i , \vp _i, t _i ) \raan
\la \D(\vR _i , \vp _i, t _i ) \raan .
\label{Eq. Factor Wigner linear response}
\end{equation}
Above we argued that the stimulus $\la \D \raan$ in the
semi-classical limit is the standard Boltzmann driving term,
Eq.\ (\ref{DrivingTermapproximation}). We will calculate the average of the
propagator $J$ perturbatively to second
order in the Chern-Simons gauge field $a$. This procedure is gauge
invariant. First notice that $\la \Delta S \raan = 0$, since $\la
\bar{a} = 0 \raan = 0$ and all averages with a $\Delta a$ at the
latest time vanishes. So, to second order in the Chern-Simons field
fluctuations is
\begin{equation}
\la J(\vR, \vp ,t; \vR _i , \vp _i, t _i ) \raan
=
\int d \vdr \int d \vdr _i \int d \rv _i  ^\prime
\int \cd \rv (\fin) \int \cd \vp (\fin) \int \cd \rv (\bi) \int \cd \vp (\bi)
\exp \left ( \frac{i}{\hbar } S_0 + \frac{i}{\hbar } \Sint \right
),
\label{Eq: PropJmidlet1}
\end{equation}
where
\begin{equation}
\Sint =
\frac{i}{2 \hbar } \la \Delta S ^2 \raan .
\end{equation}
To the same order it is
permissible in the expression for $\Delta S$ to use the approximation
$\wa(a; \vR, \vp, t) = \left . \wa(a; \vR, \vp, t) \right |_{a=0} =
w(\vp) =f(\epsilon _{\vp})$, where
$f(\epsilon)$
is the Fermi distribution function. The contribution from the
gauge field
to the effective action is then
\begin{eqnarray}
\Sint&=&
\frac{i}{2 \hbar } \la \Delta S ^2 \raan
\nonumber\\
&=&
\frac{i}{2}
\int _{t _i}^{t}ds^{\prime}\int _{t _i}^{t}ds^{\prime \prime}
\sum_{\mu^{\prime} \mu^{\prime \prime}}
\mu^{\prime}\mu^{\prime \prime}
\biggl(\matrix{-1 \matrixcomma & \dot{\rv}\left(s ^{\prime},\mu ^{\prime} \right)}\biggr)
\C\left(\rv\left(s ^{\prime},\mu ^{\prime} \right),
s ^{\prime};
\rv\left(s ^{\prime \prime},\mu ^{\prime \prime} \right),
s ^{\prime \prime} \right)
\left(\matrix{-1 \cr \dot{\rv}
\left(s ^{\prime \prime},\mu ^{\prime \prime} \right)}\right)
\nonumber\\
&&
-
\int _{t _i}^{t}ds^{\prime}\int _{t _i}^{t}ds^{\prime \prime}
\sum_{\mu^{\prime} \mu^{\prime \prime}}
\mu^{\prime}
\left\{\frac{1}{2}-w[\vp\left(s ^{\prime \prime +},\mu ^{\prime \prime}
\right)] \right\}
\biggl(\matrix{-1 \matrixcomma & \dot{\rv}\left(s ^{\prime},\mu ^{\prime} \right)}\biggr)
\bD\left(\rv\left(s ^{\prime},\mu ^{\prime} \right),
s ^{\prime};
\rv\left(s ^{\prime \prime},\mu ^{\prime \prime} \right),
s ^{\prime \prime} \right)
\left(\matrix{-1 \cr \frac{1}{m} \vp
\left(s ^{\prime \prime},\mu ^{\prime \prime} \right)}\right) . %
\nonumber\\
\label{Golchern interaction action full}
\end{eqnarray}
In the Coulomb gauge, which we consistently use below, the most singular
part of the $2\times2$ matrices $\bC$ and $\bD$ is the transverse part. Below we often only discuss
this most singular part, i.e.\ the transverse contribution to
$\Sint$,
\begin{equation}
\Sint_{\perp \perp}\left(  t_i \rightarrow t\right)   %
=
\sum_{\mu^{\prime}%
,\mu^{\prime\prime}=\pm}\mu^{\prime}\mu^{\prime\prime}\int_{t_i}^{t}ds^{\prime
}\int_{t_i}^{s^{\prime}}ds^{\prime\prime}\,{L_{\text{int}}}_{\perp\perp} 
(s^{\prime},\mu^{\prime};s^{\prime\prime},\mu^{\prime\prime})%
.%
\label{Eq Sint expres Langrangian}%
\end{equation}
For later use we have incorporated the time interval $(t_i,t)$ in which the 
interaction 
takes place into the notation for the action. The retarded Lagrangian ${L_{\text{int}}}_{\perp\perp}={L_{\text{int}}}_{\perp\perp} 
(s^{\prime},\mu^{\prime};s^{\prime\prime},\mu^{\prime\prime})$ %
for a specific pair of paths only
considering the transverse contribution is given by
\begin{eqnarray}
\frac{i}{\hbar}{L_{\text{int}}}_{\perp \perp}
&=&
\Theta\left(s^{\prime} - s^{\prime \prime} \right)
\frac{1}{\hbar}
\int \frac{d \vq}{\left(2 \pi \hbar \right)^2}
\int \frac{d \omega}{2 \pi} %
\e ^ { 
-i\omega(s^{\prime}-s^{\prime \prime})
}
\exp\left[\frac{i}{\hbar}\vq\cdot
\left(\rv\left(s ^{\prime},\mu ^{\prime} \right)-
\rv\left(s ^{\prime \prime},\mu ^{\prime \prime}
\right)\right)\right]
\nonumber\\ %
&& %
\times \frac{1}{q^2}\biggl[\vq \times \dot{\rv}\left(s ^{\prime},\mu ^{\prime}
\right)\biggr] \cdot\biggl[
\vq \times \dot{\rv}
\left(s ^{\prime \prime},\mu ^{\prime \prime} \right) \biggr]
\coth\left(\frac{\hbar \omega}{2\T}\right)
\Im D_{\perp\perp}\left(\vq ,\omega \right)
\nonumber\\
&&
-\mu ^{\prime \prime} \frac{i}{\hbar}
\Theta\left(s^{\prime} - s^{\prime \prime} \right)
\int \frac{d \vq}{\left(2 \pi \hbar \right)^2}
\int \frac{d \omega}{2 \pi}
\e ^ {
-i\omega(s^{\prime}-s^{\prime \prime}) 
}
\e^ { 
\frac{i}{\hbar}\vq\cdot
\left[\rv\left(s ^{\prime},\mu ^{\prime} \right)-
\rv\left(s ^{\prime \prime},\mu ^{\prime \prime}
\right)\right]
}
\nonumber\\
&&
\times
\frac{1}{mq^2}
\biggl[\vq \times \dot{\rv}\left(s ^{\prime},\mu ^{\prime}
\right) \biggr] \cdot
\biggl[
\vq \times \vp
\left(s ^{\prime \prime},\mu ^{\prime \prime} \right)
\biggr]
\frac{1}{2} D_{\perp \perp} \left(\vq ,\omega \right)
\left[1-2w(\vp\left(s ^{\prime \prime +},\mu ^{\prime \prime}
\right))\right]. %
\label{Eq: Golchern interaction lagrangian tran}
\end{eqnarray}
In the second term $\frac{1}{2}D_{\perp \perp} $ can be replaced
by either $\Re D_{\perp \perp}$ or
$\Im D_{\perp \perp}$ as long as we remember %
${L_{\text{int}}}$ is
retarded ($s^{\prime}>s^{\prime\prime}$).

\subsection{Recapitulation of the major results}
\label{Subsec Density matrix major results}

Summing up, the linear response of the one particle density matrix
to an external perturbation is given by Eq.\ %
(\ref{Eq. Full prop J}), which schematically can be written as
\begin{equation}
\delta w = \int \la J \; \D \raan .
\end{equation}
This expression we approximately factorize in order to arrive at
Eq.\ (\ref{Eq. Factor Wigner linear response}), or schematically
\begin{equation}
\delta w = \int \la J \raan \la \D \raan .
\end{equation}
In the semi-classical limit the stimulus $\la D \raan$ is in the classical limit given by the
driving term in the Boltzmann equation
Eq.\ (\ref{DrivingTermapproximation}).
In Eq.\ (\ref{Eq: PropJmidlet1}) the propagator $\la J 
\raan$ is represented
as a double single-particle path integral with
an action similar to the familiar Caldeira-Leggett influence
functional. Schematically we write
\begin{equation}
\la J \raan = \int \cd \text{(path $+$)} \int \cd \text{(path
$-$)}\;
\exp \left[\frac{i}{\hbar}
\left(S_0 + \Sint \right)\right].
\end{equation}
Here $S_0$ is the usual action for non-interacting particles Eq.\
(\ref{Golchern def of S0}). It does not couple the forward path $+$ %
and the backward path $-$.
$\Sint$ contains the contribution from the gauge fields and the
interactions and it couples the two paths. It is given by
Eq.\ (\ref{Golchern interaction action full}). The
imaginary part of $\Sint$ is governed by the fluctuations
Eq.\ (\ref{Def C matrix}), while the real part is determined by the
retarded gauge field propagator Eq.\ (\ref{Def D matrix}). The
distribution function also enters into the real part.


\section{Comparison with the Kubo formula}

\label{Sec KuboFormulaComparision}

\begin{figure}[t]
\centerline{\epsfig{file=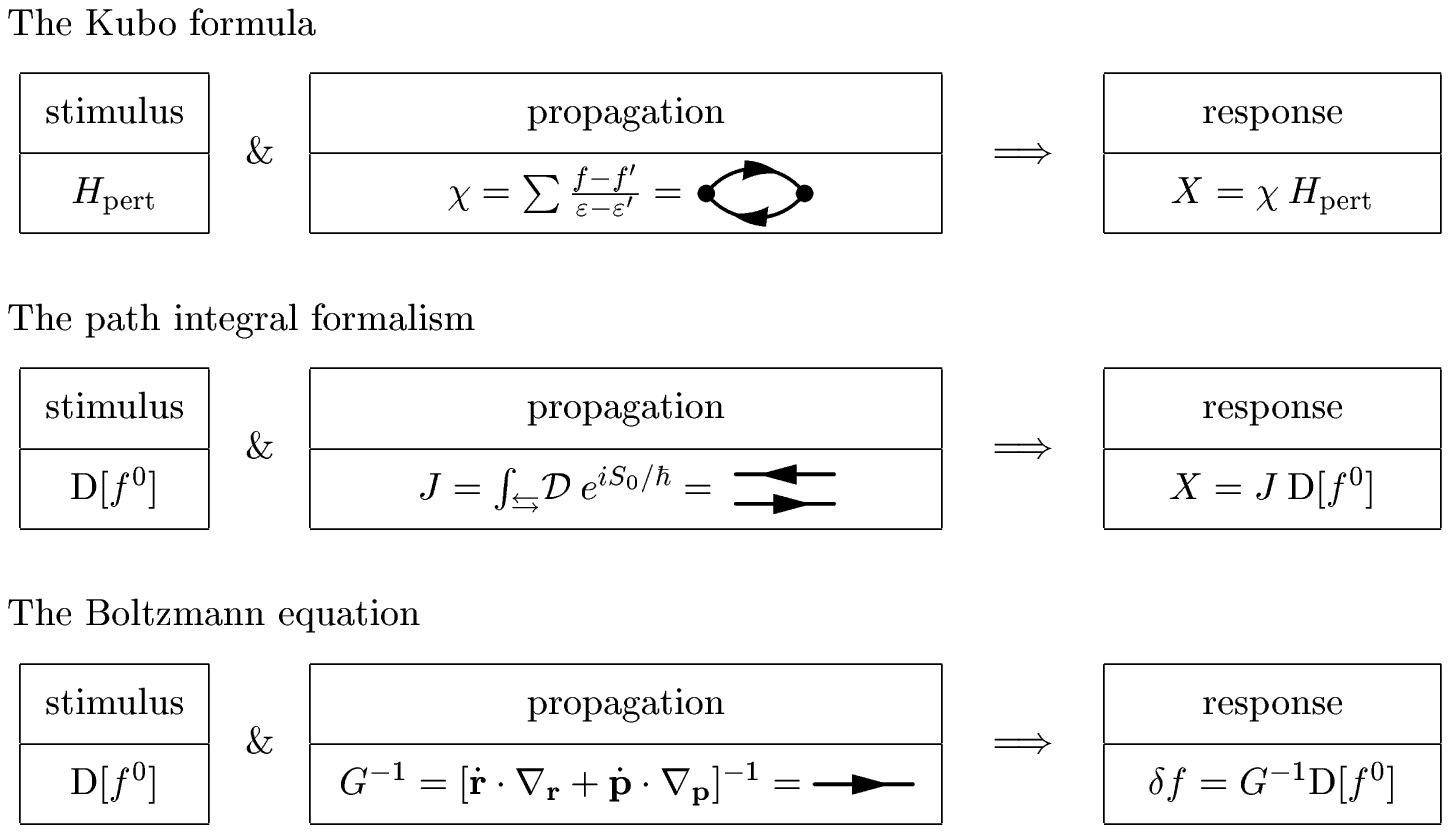, height=73mm}}
\caption{An overview of linear response theory for non-interacting particles
in the Kubo formula, the path integral formalism and the Boltzmann equation. 
The linear response results from a stimulus combined with propagation. 
The figure illustrates where the distribution function $f^0$ appears in the 
three different cases. Note in particular the absence of $f^0$ in 
the propagation part of the path integral formalism and 
the Boltzmann equation.}
\label{Fig3kubo}
\end{figure}

In equilibrium a standard way to perform linear response is to use 
the Kubo formula. Though the path integral formalism and the Kubo 
formula on a formal level are similar there are some important 
differences in the descriptions. This is most clearly illustrated by
considering a non-interacting system as is done in
Fig.\ \ref{Fig3kubo}, where we also compare to the classical 
Boltzmann equation. 
The linear response is in all three cases split into a stimulus that is 
propagated in order to give the response. In the Kubo formula the stimulus
is the applied potential, while it in the path integral formalism is the 
induced excitation, which in the semi-classical limit equals the usual 
Boltzmann stimulus Eq.\ (\ref{DrivingTermapproximation}).
Despite that
the stimulus is very different the Kubo formula and the path integral formulations
have similar pictorial representations of the propagation: 
in both cases there is a forward and backward propagating electron line.
However in the Kubo formula a line mathematically represents a Green's 
function and the propagator in the non-interacting case is the 
Lindhard function\cite{Lindhard}, 
while in the non-interacting path integral 
formalism a line is a conventional free particle Feynman path integral.
In the semi-classical limit the double Feynman path integral reduces 
to the classical path integral propagator for the Boltzmann equation
\cite{BoltzPath}.
The difference between the Kubo formula and the path integral formalism
is high lighted by the observation that the (Fermi) distribution function
in the Kubo formula is contained entirely within the propagator, while
it in the path integral formalism is the stimulus and not the propagation 
that depends 
on the distribution function for a non-interacting system.
Notice also that in the path integral formulation no statements are
made about equilibrium. 
The fact that for a non-interacting system the propagation is independent
of the actual distribution function is well known from
the Landauer-B\"{u}ttiker scattering
approach to transport through a non-interacting part of the system.

Within the Kubo formalism it is very natural to treat the
interactions perturbatively. The effect of the interaction is
traditionally divided into self-energy contributions, which alters
the propagation properties of a single line or Green's function,
and vertex corrections that connects the two lines. It is possible
to make the same distinction in a perturbative treatment of the
interactions in the path integral formulation, and we will also
use the terms self-energy and vertex corrections to distinguish
between interactions that only involves one path and those that
connect the forward and backward propagating paths.
For an interacting system the propagator in the path integral formalism
thus also depends on the actual distribution function. This is well known
in the classical limit of the Boltzmann equation, since the propagator then
depends on both the collision integral and the Landau interaction function,
but still the stimulus has already gathered the most trivial dependence
upon the distribution function.


\section{The semi-classical limit and the separation between the two paths}

\label{Sec SemiclasSeparation}

Traditionally the semi-classical limit of a double path integral
is the limit, where the important contributions are those where the
forward and backward paths are almost identical. In a discussion
of the semi-classical limit it is thus important to know how far
apart on average the forward and backward paths are. This
information is contained in the moments of the difference variable
$\Delta\rv \left(  s\right)  $. For a non-interacting system or
more generally a system with instantaneous interactions the
standard deviation of $\Delta\rv $ can be estimated from the
Heisenberg uncertainty relations between $\Delta\rv $ and its
conjugate variable, the average momentum $\vP $. That $\vP $
indeed is the conjugate variable can be read off from the
relations
\begin{eqnarray}
\frac{\partial}{\partial\vP _{\mathrm{f}}}S_{0}\left(
\vR  _{\mathrm{f}},\vP _{\mathrm{f}},t;\vR  _{\mathrm{i}%
},\vP _{\mathrm{i}},0\right)   %
& =& %
-\Delta \rv \left( t\right) \\ %
\frac{\partial}{\partial\vP _{\mathrm{i}}}S_{0}\left(
\vR  _{\mathrm{f}},\vP _{\mathrm{f}},t;\vR  _{\mathrm{i}%
},\vP _{\mathrm{i}},0\right)   %
& =&%
\Delta\rv \left(  0\right)  ,
\end{eqnarray}
where $S_{0}$ is the action for a non-interacting system, %
Eq.\ (\ref{Golchern def of S0}). 
We are mainly interested in Fermi liquids, so we prefer to use a polar
coordinate system representation of the average momenta
\begin{equation}
\vP \left(  s\right)  =P\left(  s\right)  \left( %
\matrix{ %
\cos\phi\left(  s\right) \cr %
\sin\phi\left(  s\right) } %
\right)  ,
\end{equation}
where the angle along the Fermi surface $\phi$ is measured relative to an
arbitrarily fixed $x$-axis. The conjugate variable to the size of the momentum
is
\begin{equation}
\frac{\partial}{\partial P_{\mathrm{f}}}S_{0}\left(  \vR  %
_{\mathrm{f}},\vP _{\mathrm{f}},t;\vR  _{\mathrm{i}},\vP %
_{\mathrm{i}},0\right)  =-\frac{1}{P_{\mathrm{f}}}\vP _{\mathrm{f}}%
\cdot\Delta\rv \left(  t\right)  .
\end{equation}
From the Heisenberg uncertainty relations we then expect the
standard deviation $\frac{1}{P}\vP \cdot \Delta\rv $ 
of the difference along the path
to be given by $\hbar$ divided
by the uncertainty in the
size of the momentum, i.e. 
$\frac{1}{P}\vP \cdot \Delta\rv 
\approx\hbar v_{\mathrm{F}}\left(  k_{\mathrm{B}%
}T\right)  ^{-1}$. This argument is carried through in more detail in appendix
\ref{App LongCorr} and the coefficient to $\hbar v_{\mathrm{F}}\left(
k_{\mathrm{B}}T\right)  ^{-1}$ is also shown to be $1/\sqrt{3}$.

The conjugate variable to the angle $\phi$ is
\begin{equation}
\frac{\partial}{\partial\phi_{\mathrm{f}}}S_{0}\left(  \vR  %
_{\mathrm{f}},\vP _{\mathrm{f}},t;\vR  _{\mathrm{i}},\vP %
_{\mathrm{i}},0\right)  =-\vP _{\mathrm{f}}\times\Delta\rv \left(
t\right)  .\label{Eq konjugate phi var def}%
\end{equation}
Remembering that the relevant momentum $P_{\mathrm{f}}$ approximately is the
Fermi-momentum we will denote the quantity $p_{\mathrm{F}}^{-1}\vP %
_{\mathrm{f}}\times\Delta\rv $ by the width. The Heisenberg
uncertainty relation then says that the standard deviation of the width
is approximately the Fermi wavelength $\lambda_{\mathrm{F}}$
divided by the angular uncertainty in the propagation direction.
For a well collimated beam the width is thus big, while it is small 
(of the order of $\lambda_{\mathrm{F}}$) when
measuring the current response to a homogeneous field, since in this case
the angular
uncertainty is as big as possible (of the order $1$).

The separation between the paths is not in itself directly
measurable, but it tends to show up indirectly in experiments
probing the semi-classical regime. It marks e.g.\  the cross-over
between the long wave-length regime, where the vertex
corrections are comparable to and partly cancels the self-energy and
the short wave-length regime where the vertex corrections play a minor role.
For instance in the calculation of the inverse transport time due
to impurity scattering in the famous factor $\left(  1-\cos \theta\right)$
the scattering angle $\theta$ can be
interpreted as $\frac{1}{\hbar}q\lambda
_{\mathrm{F}}  $, where $q$ is the scattering vector and,
as we saw above, $\lambda_{\mathrm{F}}$ is the width.

For the composite fermions, where the force is transmitted via a
magnetic field and thus a flux through an area, we expect the
width to mark the regime where the vertex corrections sets in. 
We can thus use the width as an effective cut-off in the
conventional calculations of the effective mass. The above
considerations regarding the width are only true for a system with
a time local action. For composite fermions this is not the case,
since the important transverse fluctuations have a characteristic
frequency around $\gc q^{2}$, see %
Eq.\ (\ref{Eq: GaugePropTransversePropInfrared}). 
We thus have to take the lowest order
correction due to the terms in the action that are not time-local into
account. We do this perturbatively.

To this end let us define the average width squared $\left(  \hbar/q_{\perp
}^{\mathrm{c}}\right)  ^{2}$ at time $s$ as
\begin{equation}
\left(  \frac{\hbar}{q_{\perp}^{\mathrm{c}}}\right)
^{2}
\equiv
\avelb p_{\mathrm{F}}^{-2}\left[ \vP
\left(  s\right) \times\Delta\rv \left(  s\right) \right]
^{2}\averb _{[0,t]},%
\label{Eq konjugate def width squared}
\end{equation}
where the average $\avell X\averl
_{[0,t]} $ is defined as
\begin{equation}
\avell X \averl_{[0,t]}  %
=\frac{\int \frac{d\vP _{\mathrm{f}}}{\left( 2\pi\hbar\right)
^{2}}\int d\vR _{\mathrm{f}}\int\frac{d\vP _{\mathrm{i}}}{\left(
2\pi \hbar\right)  ^{2}}\int d\vR _{\mathrm{i}}\,\calR \left( \vP
_{\mathrm{f}},\vR _{\mathrm{f}}\right) \D \left( \vP
_{\mathrm{i}},\vR _{\mathrm{i}}\right) 
\pilf X\exp\left[ \frac{i}{\hbar}S_{0}\left(
\vR  _{\mathrm{f}},\vP _{\mathrm{f}},t;\vR  %
_{\mathrm{i}},\vP _{\mathrm{i}},0\right)  \right]  \pirf _{[0,t]}}{\int\frac{d\vP %
_{\mathrm{f}}}{\left(  2\pi\hbar\right)  ^{2}}\int d\vR  _{\mathrm{f}%
}\int\frac{d\vP _{\mathrm{i}}}{\left(  2\pi\hbar\right)  ^{2}}\int
d\vR  _{\mathrm{i}}\,\calR \left(  \vP _{\mathrm{f}%
},\vR  _{\mathrm{f}}\right)  \D \left(  \vP _{\mathrm{i}%
},\vR  _{\mathrm{i}}\right)  \pilf
\exp\left[  \frac{i}{\hbar}S_{0}\left(  \vR  _{\mathrm{f}%
},\vP _{\mathrm{f}},t;\vR  _{\mathrm{i}},\vP _{\mathrm{i}%
},0\right)  \right]  \pirf _{[0,t]}} .
\label{Eq konjugate def <<>>t}%
\end{equation}
Here $\pill X \pirl _{[0,t]}$ is a short hand notation 
for the path integral over paths starting at time $0$ and ending at time 
$t$ with the Wigner function variables as boundary conditions. 
$\pill X \pirl _{[0,t]}$ is defined such as to include the 
action $S_{\mathrm{int}}(t,0)$ arising from the interactions 
within the interval $[0,t]$ 
\begin{equation}
\pill X \pirl_{[0,t]}  %
=
\int_{[0,t]}\cd   %
\vR  \int_{[0,t]}\cd   \vP \int_{[0,t]}\cd
\Delta\vp \int_{[0,t]}\cd   \Delta \rv 
\e^ {
\frac{i}{\hbar}S_{\mathrm{int}}(t,0) 
}
X.
\label{Eq konjugate def <<<>>>t}%
\end{equation} 
When no confusion is possible we will the index $[0,t]$.
In the semi-classical limit the stimulus $\D \left(  \vP _{\mathrm{i}},\vR  %
_{\mathrm{i}}\right)  $ is given by
Eq.\ (\ref{DrivingTermapproximation}). For a current measurement
the response $\calR \left(  \vP _{\mathrm{f}},\vR
_{\mathrm{f}}\right)
=-e\vv _{\mathrm{f}}\Omega^{-1}=-e\frac{1}{m}\vp _{\mathrm{f}%
}\Omega^{-1}$. $\Omega$ is the volume of the sample. The time $t$ is the
time-interval during which the paths evolve. Below we will show that in the
long time limit the cut-off is independent of $t$ and $s$. We only consider
this limit below and for our purpose $t$ can be taken to infinity or rather be
put equal to a scattering time. Notice that according to 
Eq.\ (\ref{Eq: PropJmidlet1})
the last factor in the denominator is
\begin{equation}
\pilB
\exp\left[  \frac{i}{\hbar}S_{0}\left(
t,\vR  _{\mathrm{f}},\vP _{\mathrm{f}},t;
\vR  _{\mathrm{i}},\vP _{\mathrm{i}},0\right)  \right]  
\pirB _{\indexnegspacepirB [0,t]}=\la J\left( \vR %
_{\mathrm{f}},\vP  _{\mathrm{f}},t;\vR _{\mathrm{i}},\vP  %
_{\mathrm{i}},0\right) \raa .
\end{equation}
It could easily be argued that it is better to use the absolute value of both
the response $\calR \left(  \vP _{\mathrm{f}},\vR  %
_{\mathrm{f}}\right)  $ and the stimulus $\D \left(  \vP %
_{\mathrm{i}},\vR  _{\mathrm{i}}\right)  $ in the definition of
$\avell X \averl $, Eq.\ %
(\ref{Eq konjugate def <<>>t}), since the response might be the sum 
of a positive contribution from
paths originating in one part of the stimulus phase space and a
negative contribution from other parts of the phase space. The
width squared of the positively contributing paths taken as a
group is then canceled by the
width of the negatively contributing paths taken as a group, the
overall result might even be a negative width squared. This does
not seem to be terribly important in the examples discussed below
so for simplicity we stick with the above definition.

We next use the conjugate variable relationship Eq.\ %
(\ref{Eq konjugate phi var def}) to perform the rewriting
\begin{equation}
p_{\mathrm{F}}^{-2}\left[  \vP \left(  s\right)  \times\Delta
\rv \left(  s\right)  \right]  ^{2}\e ^ { 
\frac{i}{\hbar}%
S_{0}\left[  \vR  _{\mathrm{f}},\vP _{\mathrm{f}%
},t;\vR  _{\mathrm{i}},\vP _{\mathrm{i}},0\right]  
}=
\frac{\partial}{\partial\phi\left(  s\right)  }\e ^ { 
\frac{i}{\hbar }S_{0}\left[ \vR  \left(  s\right)
,\vP \left( s\right),s ;\vR _{\mathrm{i}},\vP _{\mathrm{i}},0\right]
} 
\frac{\partial}{\partial\phi\left(  s\right)
}\e^ { 
\frac{i}{\hbar}S_{0}\left[ \vR  _{\mathrm{f}},\vP _{\mathrm{f}%
},t;\vR  \left(  s\right)  ,\vP \left(  s\right),s  \right] 
}
\end{equation}
inside the definition of the width squared Eq.\ %
(\ref{Eq konjugate def width squared}). The action due to the
interactions $S_{\mathrm{int}}$ is split into three parts
\begin{eqnarray}
S_{\mathrm{int}}\left( t,0\right)   %
& =&%
\sum_{\mu^{\prime}%
,\mu^{\prime\prime}=\pm}\mu^{\prime}\mu^{\prime\prime}\int_{0}^{t}ds^{\prime
}\int_{0}^{s^{\prime}}ds^{\prime\prime}\,L_{\mathrm{int}}\\ %
&=&  %
S_{\mathrm{int}}\left(  s,0\right)
+S_{\mathrm{int}}\left( t,s\right)
+\sum_{\mu^{\prime},\mu^{\prime\prime}=\pm}\mu^{\prime
}\mu^{\prime\prime}\int_{s}^{t}ds^{\prime}\int_{0}^{s}ds^{\prime\prime
}\,L_{\mathrm{int}},%
\label{Eq konjugate Sint 3 terms}%
\end{eqnarray}
where the Lagrangian $L_{\mathrm{int}}$ only considering the
transverse contribution is given by Eq.\ %
(\ref{Eq: Golchern interaction lagrangian tran}).
In total 
\begin{eqnarray}
\lefteqn{
\left(  \frac{\hbar}{q_{\perp}^{\mathrm{c}}}\right)  ^{2}  %
=
\lambda_{\mathrm{F}}^{2}
\calN %
\int\frac{d\vP _{\mathrm{f}}}{\left(  2\pi\hbar\right) ^{2}}\int
d\vR _{\mathrm{f}}\int\frac{d\vP _{\mathrm{i}}}{\left(  2\pi
\hbar\right)  ^{2}}\int d\vR  _{\mathrm{i}}\,\calR \left( \vP
_{\mathrm{f}},\vR  _{\mathrm{f}}\right) \D \left(
\vP _{\mathrm{i}},\vR  _{\mathrm{i}}\right)   
\int\frac{d\vP \left(  s\right) }{\left(
2\pi\hbar\right)  ^{2}}\int d\vR  \left(  s\right)
}
\nonumber\\%
&&
\times
\pilBb \pilBb \e ^ { 
\frac{i}{\hbar}\sum_{\mu^{\prime},\mu^{\prime\prime
}=\pm}\mu^{\prime}\mu^{\prime\prime}\int_{s}^{t}ds^{\prime}\int_{0}%
^{s}ds^{\prime\prime}\,L_{\mathrm{int}}
}
\frac{\partial}{\partial\phi\left(  s\right)  }\e ^ { 
\frac{i}{\hbar }S_{0}\left[  \vR  \left( s\right)
,\vP \left( s\right),s ;\vR _{\mathrm{i}},\vP _{\mathrm{i}},0\right]
}
\frac{\partial}{\partial \phi\left(  s\right)  }\e ^ { 
\frac{i}{\hbar}S_{0}\left[ \vR _{\mathrm{f}},\vP
_{\mathrm{f}},t;\vR  \left( s\right) ,\vP \left(  s\right),s \right]
} \pirBb _{\indexnegspacepirBb [0,s]}\pirBb _{\indexnegspacepirBb [s,t]},
\end{eqnarray}
where the normalization is contained inside the new constant $\calN$,
\begin{equation}
\calN^{-1} =
\int\frac{d\vP _{\mathrm{f}}}{\left( 2\pi\hbar\right) ^{2}}\int
d\vR _{\mathrm{f}}\int\frac{d\vP _{\mathrm{i}}}{\left( 2\pi
\hbar\right)  ^{2}}\int d\vR  _{\mathrm{i}}\,\calR \left( \vP
_{\mathrm{f}},\vR  _{\mathrm{f}}\right) \D \left(
\vP _{\mathrm{i}},\vR  _{\mathrm{i}}\right)  
\pill \exp\left[
\frac{i}{\hbar}S_{0}\left(
\vR  _{\mathrm{f}},\vP _{\mathrm{f}},t;\vR  %
_{\mathrm{i}},\vP _{\mathrm{i}},0\right)  \right]  \pirl _{[0,t]}%
.
\end{equation}

The contribution due to the interaction is treated perturbatively; in
particular we perform the expansion in the part that contains the scattering
events that begin before the observation time $s$ and finish later than $s$:
$\sum_{\mu^{\prime},\mu^{\prime\prime}=\pm}\mu^{\prime}\mu^{\prime\prime}%
\int_{s}^{t}ds^{\prime}\int_{0}^{s}ds^{\prime\prime}\,L_{\mathrm{int}}$.

\subsection{Zeroth order or the time-local limit revisited}

To lowest order or in the so-called time-local approximation, the
last term in Eq.\ (\ref{Eq konjugate Sint 3 terms}) vanishes and
the width squared is
\begin{eqnarray}
\left(  \frac{\hbar}{q_{\perp}^{\mathrm{c}}}\right)  ^{2}  %
&= & %
\lambda_{\mathrm{F}}^{2}
\calN \int\frac{d\vP _{\mathrm{f}}}{\left(  2\pi
\hbar\right)  ^{2}}\int d\vR  _{\mathrm{f}}\int\frac{d\vP %
_{\mathrm{i}}}{\left(  2\pi\hbar\right)  ^{2}}\int d\vR  _{\mathrm{i}%
}\,\calR \left( \vP _{\mathrm{f}},\vR _{\mathrm{f}}\right) \D
\left(
\vP _{\mathrm{i}},\vR  _{\mathrm{i}}\right) 
\int\frac{d\vP \left( s\right) }{\left(
2\pi\hbar\right)  ^{2}}\int d\vR  \left( s\right)
\nonumber\\ %
&& %
\times  \frac{\partial
}{\partial\phi\left(  s\right)  }\la J\left[ \vR
\left(
s\right)  ,\vP \left(  s\right),s  ;\vR  _{\mathrm{i}},\vP %
_{\mathrm{i}},0\right]  \raa 
\frac{\partial}{\partial\phi\left(  s\right)  } 
\la J\left[
\vR _{\mathrm{f}},\vP _{\mathrm{f}},t;\vR \left(
s\right) ,\vP \left(  s\right),s  \right] \raa .
\end{eqnarray}
Let us study two distinct examples. The first is the case of
homogeneous
driving, where $\D \left(  \vP _{\mathrm{i}},\vR  %
_{\mathrm{i}}\right)  =\D \left(  \vP _{\mathrm{i}}\right) $. We
furthermore assume the system is rotation-invariant such that
\begin{equation}
\int d\vR  _{\mathrm{i}}\,\la J\left[  \vR  \left(
s\right)  ,\vP \left(  s\right),s  ;\vR  _{\mathrm{i}},\vP %
_{\mathrm{i}},0\right]\raa
\end{equation}
only depends on the size of the momenta $P_{\mathrm{i}}$, $P\left(
s\right) $ and the relative momentum angle $\phi\left(  s\right)
-\phi_{\mathrm{i}}$. In this case it is nice to expand the angular
dependence of the stimulus on the eigenfunctions 
$\exp\left(  i\nu\phi_{\mathrm{i}}\right)  $ %
for the angular
momentum operator in the $z$-direction
\begin{equation}
\D \left(  \vP _{\mathrm{i}}\right)  =\sum_{\nu}\exp\left(
i\nu\phi_{\mathrm{i}}\right)  \D _{\nu}\left(
P_{\mathrm{i}}\right)  .
\end{equation}
The width squared from the $\nu$'th mode is then
$\nu^{2}\lambda_{\mathrm{F}}^{2}$. In a path integral calculation of
the impurity scattering integral in the Boltzmann equation the well known 
factor $1- \cos(\nu \theta)$ factor originates from an average of 
$2 \sin^2\left(\frac{1}{2}\theta \frac{\partial}{\partial \phi(s)}\right)$,
so as expected it is the width $\nu\lambda_{\mathrm{F}}$ which sets 
the boundary between the small angle scattering regime, where 
the vertex corrections given by $\cos(\nu \theta)$) are important and a 
large scattering regime, where the vertex correction average out.
Notice that for an elastic scattering system this averaging has to be 
performed via the boundary conditions, while in general 
an interacting system is self-averaging.

As another example we consider the gedanken experiment typically considered
in connection with the interpretation of a single-particle Greens function.
A particle is fed into the system at the Fermi surface and in a direction 
parallel to the $x$-axis.
In order to make it a bit more realistic we give the direction a finite 
spread $\Delta_{\mathrm{i}}$ i.e.\ we take 
$\calR \propto \exp\left(\frac{1}{2\Delta_{\mathrm{i}}^2} 
\phi_{\mathrm{i}}^2\right)$. A time $t$ later we measure the probability the
particle is still moving along the $x$-axis with a directional spread of 
$\Delta_{\mathrm{f}}$ i.e.\ we take 
$\D \propto \exp\left(\frac{1}{2\Delta_{\mathrm{f}}^2} 
\phi_{\mathrm{f}}^2\right)$. The width squared is in this case 
$\lambda_{\mathrm{F}}^2\left(\frac{1}{\Delta_{\mathrm{i}}^2}+
\frac{1}{\Delta_{\mathrm{f}}^2}\right)$. In particular the width diverges as 
the angular spread to the power $-1$.

\subsection{The first order non time-local correction }

Let us move on to discuss the first order correction to the width squared due
to the non time-local part of the interaction $\sum_{\mu^{\prime},\mu
^{\prime\prime}=\pm}\mu^{\prime}\mu^{\prime\prime}\int_{s}^{t}ds^{\prime}%
\int_{0}^{s}ds^{\prime\prime}\frac{i}{\hbar}\,L_{\mathrm{int}}$
\begin{eqnarray}
\lefteqn{
\Delta\left(  \frac{\hbar}{q_{\perp}^{\mathrm{c}}}\right)  ^{2}  %
=
\lambda_{\mathrm{F}}^{2}
\calN \frac{i}{\hbar}\int\frac{d\vP _{\mathrm{f}}}{\left( 2\pi\hbar\right)
^{2}}\int d\vR _{\mathrm{f}}\int\frac{d\vP _{\mathrm{i}}}{\left(
2\pi\hbar\right)  ^{2}}\int d\vR  _{\mathrm{i}}\,\calR \left( \vP
_{\mathrm{f}},\vR  _{\mathrm{f}}\right) \D \left(
\vP _{\mathrm{i}},\vR  _{\mathrm{i}}\right) 
\int\frac{d\vP \left(
s\right)  }{\left(  2\pi\hbar\right)  ^{2}}\int d\vR  \left(
s\right)
}\nonumber\\
&& %
\times
\sum_{\mu^{\prime},\mu^{\prime\prime}=\pm}\mu^{\prime}\mu^{\prime\prime}%
\int_{s}^{t}ds^{\prime}\int_{0}^{s}ds^{\prime\prime} 
\pilB \pilB L_{\mathrm{int}}(s^{\prime},\mu^{\prime};s^{\prime \prime},\mu^{\prime \prime})
\frac{\partial}{\partial\phi\left(
s\right) }\e ^ { 
\frac{i}{\hbar}S_{0}\left[ \vR  \left(
s\right)  ,\vP \left(  s\right),s  ;\vR  _{\mathrm{i}},\vP %
_{\mathrm{i}},0\right]  
}
\frac{\partial}{\partial \phi\left( s\right)  }\e^{ 
\frac{i}{\hbar}S_{0}\left[ \vR _{\mathrm{f}},\vP
_{\mathrm{f}},t;\vR  \left( s\right) ,\vP \left(  s\right),s \right]
}\pirB _{\indexnegspacepirB [0,s]}\pirB
_{\indexnegspacepirB [s,t]}.%
\nonumber\\
\label{Eq konjugate Firstorder 1}%
\end{eqnarray}
We have here neglected the contribution from the expansion of the
normalization $\calN $, since it turns out to cancel a term in the
expansion of Eq. (\ref{Eq konjugate Firstorder 1}) we anyway
neglect in our discussion below. Let us here specialize to the
contribution from the long wave-length fluctuations, where $\vq
\cdot\Delta\rv \ll1$, i.e.\
\begin{eqnarray}
\frac{i}{\hbar}\,L_{\mathrm{int}} 
(s^{\prime},\mu^{\prime};s^{\prime\prime},\mu^{\prime\prime}) %
& =&%
\frac{1}{\hbar}\int\frac{d\vq %
}{\left(  2\pi\hbar\right)
^{2}}\int\frac{d\omega}{2\pi}\coth\left(
\frac{\hbar\omega}{2k_{\mathrm{B}}T}\right) \Im D_{\perp
\perp}\left( \vq ,\omega\right) \left[\frac{1}{q}\vq \times\dot{\vR}
\left( s^{\prime}\right)\right]\cdot\left[ \frac{1}{q}\vq \times\dot{\vR}  \left(
s^{\prime\prime}\right)\right]    \nonumber\\ %
&& %
\times
\e ^ {
-i\omega\left( s^{\prime}-s^{\prime \prime}\right)
}\e^{
\frac{i}{\hbar}\vq \cdot\left[ \vR \left(
s^{\prime}\right)  -\vR  \left( s^{\prime\prime}\right) \right]
+ 
\frac{i}{\hbar}\frac{1}{2}\vq \cdot\left[
\mu^{\prime}\Delta\rv \left(  s^{\prime}\right) -\mu^{\prime\prime
}\Delta\rv \left( s^{\prime\prime}\right) \right]  
}.
\end{eqnarray}
To lowest order in the interaction the extra contribution from
$L_{\mathrm{int}} $ to the width squared is in the long wave-length limit
\begin{eqnarray}
\Delta\left(  \frac{\hbar}{q_{\perp}^{\mathrm{c}}}\right)  ^{2}  %
&=& %
\calN \int\frac{d\vP _{\mathrm{f}}}{\left( 2\pi\hbar\right)
^{2}}\int d\vR _{\mathrm{f}}\int\frac{d\vP _{\mathrm{i}}}{\left(
2\pi\hbar\right)  ^{2}}\int d\vR  _{\mathrm{i}}\,\calR \left( \vP
_{\mathrm{f}},\vR  _{\mathrm{f}}\right) \D \left(
\vP _{\mathrm{i}},\vR  _{\mathrm{i}}\right) 
\frac{1}{p_{\mathrm{F}}^{2}}\int\frac{d\vP \left( s\right)
}{\left( 2\pi\hbar\right)  ^{2}}\int d\vR  \left( s\right)
\nonumber\\ %
&& %
\times
\frac{1}{\hbar}\int_{s}^{t}ds^{\prime}\int_{0}^{s}ds^{\prime\prime}\int
\frac{d\vq }{\left(  2\pi\hbar\right) ^{2}}\int\frac{d\omega}{2\pi
}\coth\left(  \frac{\hbar\omega}{2k_{\mathrm{B}}T}\right)  \Im %
D_{\perp \perp}\left(  \vq,\omega\right)  \e^{
-i\omega\left(
s^{\prime }-s^{\prime\prime}\right)  
}
\nonumber\\ %
&& %
\times
\pilBb \pilBb 
\frac{1}{\hbar}\vq \cdot\Delta\rv \left( s^{\prime}\right)
\left[
\frac{1}{q}\vq \times\dot{\vR}  \left(
s^{\prime}\right) %
\right]
\cdot
\left[
\frac{1}{q}\vq \times\dot{\vR}  \left(
s^{\prime\prime}\right) %
\right]
\frac{1}{\hbar}\vq \cdot\Delta\rv\left( s^{\prime}\right)
\nonumber\\ %
&& %
\times
\frac{\partial}{\partial\phi\left(  s\right) }\e^{ 
\frac{i}{\hbar }S_{0}\left[  \vR  \left(  s\right)
,\vP \left( s\right),s ;\vR _{\mathrm{i}},\vP _{\mathrm{i}},0\right]
+\frac
{i}{\hbar}\frac{\partial\varepsilon_{\vp \left(  s\right)  }}%
{\partial\vp \left(  s\right)  }\cdot\vq \left( s-s^{\prime
\prime}\right)  
}  
\frac{\partial}{\partial\phi\left(  s\right)  }\e^{ 
\frac{i}{\hbar
}S_{0}\left[  \vR  _{\mathrm{f}},\vP _{\mathrm{f}%
},t;\vR  \left(  s\right)  ,\vP \left(  s\right),s  \right] +\frac
{i}{\hbar}\frac{\partial\varepsilon_{\vp \left(  s\right)  }}%
{\partial\vp \left(  s\right)  }\cdot\vq \left(
s^{\prime }-s\right)  
}  
\pirBb _{\indexnegspacepirBb [0,s]}\pirBb _{\indexnegspacepirBb [s,t]}.
\nonumber\\
\end{eqnarray}
To lowest order we will neglect the effect of
$S_{\mathrm{int}}\left( s,0\right)  $ and
$S_{\mathrm{int}}\left(  t,s\right) $ inside the path integrals 
$\pill\pirl _{\indexnegspacepirBb [0,s]} $ and
$\pill\pirl _{\indexnegspacepirBb [s,t]} $ respectively. The product
$\Delta\rv \left(  s^{\prime}\right)  \Delta \rv \left(
s^{\prime\prime}\right)  $ we will replace by its average value.
The most important contribution comes from the longitudinal part
of $\Delta\rv \left(  s^{\prime}\right) \Delta\rv \left(
s^{\prime\prime}\right)  $. Furthermore it turns out that the
relevant time-scale in the above integral $\hbar\left(
v_{\mathrm{F}}q\right)  ^{-1}$ for the important fluctuations is
smaller than the correlation time of $\Delta\rv \left(
s^{\prime}\right)  \Delta\rv \left(
s^{\prime\prime}\right)  $, which is of the order $\hbar\left(  k_{\mathrm{B}%
}T\right)  ^{-1}\frac{\hbar \gc p_{\mathrm{F}}^{2}}{v_{\mathrm{F}}p_{\mathrm{F}}%
}\approx\hbar\left(  k_{\mathrm{B}}T\right)  ^{-1}$, so we replace
$\Delta\rv \left(  s^{\prime}\right)  \Delta\rv \left(
s^{\prime\prime}\right)  $ by its equal time longitudinal average
value $\frac{1}{3}\frac{\hbar^{2}v_{\mathrm{F}}^{2}}{\left(
k_{\mathrm{B}}T\right)
^{2}}$, see appendix \ref{App LongCorr}. %
Within these approximations the correction to the width squared is
\begin{equation}
\Delta\left(  \frac{\hbar}{q_{\perp}^{\mathrm{c}}}\right)  ^{2}  %
=
\frac{1}{6}\frac{\hbar^{2}v_{\mathrm{F}}^{2}}{\left(  k_{\mathrm{B}%
}T\right)  ^{2}}\frac{1}{p_{\mathrm{F}}^{2}}\int_{0}^{\infty}\frac{dq}%
{2\pi\hbar}q\int\frac{d\omega}{2\pi}\coth\left(  \frac{\hbar\omega
}{2k_{\mathrm{B}}T}\right)  \left[  -\Im D_{\perp
\perp} %
\left(
\vq,\omega\right)  \right]  .%
\end{equation}
For the composite fermions the spectral function %
$-\Im D_{\perp \perp}\left( \vq ,\omega\right)
\approx2\pi\hbar\frac{q}{p_{\mathrm{F}}}\frac
{\omega}{\omega^{2}+\left(  \g_{\eta}q^{1+\eta}\right)  ^{2}}$, %
Eq.\ (\ref{Eq: GaugePropTransversePropInfrared}), so the width
squared is approximately
\begin{equation}
\frac{\hbar^{2}}{q_{\mathrm{c}\perp}^{2}}\approx\lambda_{\mathrm{F}}%
^{2}\left[  1+\left(  \frac{v_{\mathrm{F}}p_{\mathrm{F}}}{2k_{\mathrm{B}}%
T}\right)  ^{(2\eta-1)/(1+\eta)}\left(  \frac{v_{\mathrm{F}}p_{\mathrm{F}}%
}{2\hbar \gc p_{\mathrm{F}}^{\eta+1}}\right)  ^{3/(1+\eta)}2^{3/(1+\eta)}\frac
{1}{9}\csc\frac{3\pi}{2\left(  1+\eta\right)  }\right]  .
\label{Eq SemiClasSep Final general}
\end{equation}
Here $\eta=1$ corresponds to our prime example, the Coulomb interaction
\begin{equation}
\frac{\hbar^{2}}{q_{\mathrm{c}\perp}^{2}}\approx\left(\frac{\hbar
v_{\mathrm{F}}}{2k_{\mathrm{B}}T}\lambda_{\mathrm{F}}^{3}\right)^{1/2}
\left(
\frac{v_{\mathrm{F}}p_{\mathrm{F}}}{2 \coulener} \right)
^{3/2}\frac{4}{9}\left(\frac{4}{\pi}\right)^{3/4}. %
\label{Eq widthcutoffsquare}%
\end{equation}

We
expect the Fermi energy $\frac{1}{2}v_{\mathrm{F}}p_{\mathrm{F}}$ to be of the
same order of magnitude as the energy scale set by the interaction i.e.\
the average Coulomb energy $\coulener = \frac{e^2}{4 \pi \permittivity} \sqrt{n}$. 
The factor $\left(
\frac{v_{\mathrm{F}}p_{\mathrm{F}}}{2 \coulener }\right)
^{3/2}\frac{4}{9}\left(\frac{4}{\pi}\right)^{3/4}$ is
thus of the order $1$ and the width is approximately given by the geometric
mean of the Fermi wave-length and the de Broglie wave-length.

In perturbative calculations of the self-energy the cut-off is
often formulated in terms of a frequency cut-off. The frequency cut-off
corresponding to Eq.\ (\ref{Eq widthcutoffsquare}) is
\begin{equation}
\omega_{\mathrm{cut}}
=\gc q_{\mathrm{c}\perp}^{2}\approx %
\hbar^{-1}\sqrt{\coulener\T} \,\,
\frac{9}{4}\left(\frac{\pi}{4}\right)^{5/4}\left(
\frac{2 \coulener}{v_{\mathrm{F}}p_{\mathrm{F}}}\right)
^{2}.
\end{equation}
We want to emphasize that this transverse cut-off is considerably larger than
the longitudinal cut-off, which in appendix \ref{App LongCorr} is found to be 
$\sqrt{3}%
\hbar^{-1}k_{\mathrm{B}}T$. In the next section we show that 
in case of Coulomb interaction
the effective mass depends logarithmically on $\omega_{\mathrm{cut}}$, so 
the temperature dependence of $\omega_{\mathrm{cut}}$ and thus the inverse 
width squared is $\sqrt{T} |\log T|^2$, which tends much slower to zero 
than the more conventional linear temperature scaling \cite{LeeBoltz}.


\section{The effective mass}

\label{Sec effectivemass}

In the last section we calculated the width between the forward and backward
path $\frac{\hbar}{q_{\mathrm{c}\perp}}$,
Eq.\ (\ref{Eq widthcutoffsquare}), and we hinted at how this width might 
play the role of an effective
cut-off in the calculation of the effective mass. In this section we
elaborate a bit more on this connection.

Experimentally the effective mass of a particle is determined by
simultaneously measuring the momentum and the velocity. In a Fermi liquid the
momentum is typically the Fermi momentum. The velocity measurement typically
involves measuring how far the particle has propagated in a known time-span
$t$. This is a little bit at odds with the semi-classical limit, where the
external fields are considered to be smooth, so it is hard to define, yet
measure, distances shorter than a transport mean free path length. This is
necessary in a mass measurement, since the time $t$ must be 
comparable to or smaller than the transport time to make 
the path reasonably well
defined. Having said this one might hope for an intermediate regime, where it is
possible to at least put some limits on the effective mass, while still
maintaining the semi-classical line of thought.

One attempt at this is done by Willett, West, %
and Pfeiffer \cite{WillettEksperiment}, who uses a surface
acoustic wave (SAW) to generate an external electric field with a
wave-length comparable to the mean free path. They then apply an
external magnetic field and looks for resonances between the
cyclotron radius and the wave-length of the SAW field. In order to
observe these resonances the cyclotron frequency cannot be orders
of magnitude larger than frequency of the SAW wave \cite{Wølfe}.
This leads to an upper estimate of the elapsed time and thus the
effective mass.

We will denote this sort of experiment as a semi-classical mass determination,
since the forward and backward paths together perform the cyclotron motion.
This is in contrast to experiments probing the single-particle energy
spectrum, like Shubnikov-de Haas measurements. In such experiments according 
to the Bohr-Sommerfeld quantization scheme it is the phase of a single
path that causes the resonance.

Below we show that in the semi-classical experiments the width
between the two paths or in other words the enclosed area acts as
an effective cut-off on the contribution from the magnetic field fluctuations
to the effective mass. In the Bohr-Sommerfeld
resonance scheme the cut-off is supplied by the area enclosed by
the path itself \cite{Mucciolo}. This area is much bigger than the
semi-classical area and the cut-off is correspondingly smaller.

In order to extract the semi-classical mass from the path integral
description it is in principle necessary to analyze the concrete
experiment in detail. Luckily it turns out that in the semi-classical limit it 
is possible to come
up with an approximately local relation between the velocity and the
momentum. This implies the effective mass is almost independent of 
the concrete mass measuring experiment in the semi-classical limit.

For a non-interacting system the derivative of the action with respect to the
momentum yields the Hamilton equation for the velocity $\frac{\delta S_{0}%
}{\delta p\left(  s\right)  }=\dot{r}-\frac{\partial H}{\partial
p}=0$. For the two-path action it is the derivative with respect
to the difference momentum $\Delta\vp \left(  s\right)  $
\begin{eqnarray}
\frac{\partial S}{\partial\Delta\vp \left(  s\right)  }  &
=& %
\dot{\vR}  \left(  s\right)  -\frac{1}{m}\vP \left( s\right)
-\int_{s}^{t}ds^{\prime}\int\frac{d\vq }{\left(  2\pi\hbar\right) ^{2}%
}\int\frac{d\omega}{2\pi}\,D_{\perp\perp}\left(
\vq ,\omega\right) 
\exp\left\{ \frac{i}{\hbar}\vq \cdot\left[ \vR  \left(
s^{\prime}\right)  -\vR  \left(  s\right) \right]  \right\}
\exp\left[ -i\omega\left(  s^{\prime}-s\right) \right]
\nonumber\\ %
&& %
\times \sum_{\mu \,,\mu^{\prime}=\pm}
\mu \mu^{\prime}
\e^{ 
\frac{i}{\hbar}\mu\frac{1}{2}\vq %
\cdot\left[\Delta\rv \left(s^{\prime}\right)  -\Delta\rv \left(  s\right) 
\right]
} 
\,\frac{1}{q}\vq \times\left[ \dot{\vR}
\left( s^{\prime }\right) +\frac{1}{2}\mu^{\prime}\Delta\dot{\rv}
\left(  s^{\prime
}\right)  \right] \nonumber\\ %
&& %
\times
\frac{\partial}{\partial\vP }\,\left\{ \frac{1}{2}-w\left[ \vP
\left(  s^{+}\right) +\mu\frac{1}{2}\Delta\vp \left( s^{+}\right)
\right] \frac{1}{mq}\vq \times\left[ \vP \left( s\right)
+\mu\frac{1}{2}\Delta\vp \left( s\right) \right]  \right\}
.\label{Eff mass Eq: Action derivative 1}%
\end{eqnarray}
that yields the relation between the velocity and the momentum, at
least in a stationary phase approximation where $\frac{\partial
S}{\partial \Delta\vp \left(  s\right)  }=0$. On average the
stationary phase condition is exact, i.e.\
\begin{equation}
\avelBb \frac{\partial S}{\partial\Delta\vp
\left( s\right)  }\averBb =0 .
\end{equation}
Notice that
after performing the integral over
$\vR  \left(  s\right)  $ and $\Delta\rv \left( s\right)  $ inside
the average as defined in Eq.\ (\ref{Eq konjugate def <<>>t}) we have 
$\Delta\vp \left(  s^{+}\right)  =\Delta\vp %
\left(  s\right)  -\vq $ and $\vP \left(  s^{+}\right) =\vP \left(
s\right)  -\frac{1}{2}\mu\vq $. The
terms with $\mu=\mu^{\prime}$ in Eq.\ (\ref{Eff mass Eq: Action
derivative 1}) i.e.\ those terms where the interaction connect
either the forward or backward path with itself gives rise to that
part of the renormalization of the effective mass that is
attributed to the self-energy in the usual perturbative
treatments. This renormalization mainly depends on the local
properties of a single path and it is thus independent of which
pair of paths is relevant in the given experiment. On the other
hand the terms with $\mu\neq\mu^{\prime}$ in %
Eq.\ (\ref{Eff mass Eq: Action derivative 1}), which are
associated with the vertex corrections, depend on both the forward
and the backward paths. The renormalization due to these terms
thus depends on the exact experiment. For instance in an
interference experiment the two paths are almost independent and
the vertex contribution is expected to vanish due to destructive
interference in the phases associated with each path individually.
In the semi-classical limit, where the forward and backward paths
typically are close to each other it is fortunately possible to
derive a universal expression for the renormalization of the
mass. Below it will be shown the interaction modes with a
wave-length longer than the separation between the paths 
does not renormalize the mass, while the renormalization due to the 
interaction modes with a shorter wave-length is given by the
usual self-energy contribution. 
A possible interpretation is that the particle cannot
distinguish between the field from a long wave-length interaction
and an external potential.

For simplicity we neglect the extra terms arising from the
velocity dependence of the force, i.e.\ we treat the interaction as
scalar. With this
small neglect the average of the velocity, denoted $\vv _{\mathrm{F}%
}=\avelb \dot{\vR}  \averb
$, is
\begin{eqnarray}
\vv _{\mathrm{F}}  & = &
\frac{1}{m}\vp _{\mathrm{F}}-\int_{s}%
^{t}ds^{\prime}\int\frac{d\vq }{\left(  2\pi\hbar\right) ^{2}}\int
\frac{d\omega}{2\pi}v_{\mathrm{F}}m^{-1}p_{\mathrm{F}}D_{\perp\perp}\left(
\vq ,\omega\right)  \e^ {
-i\omega\left( s^{\prime}-s\right) 
}  
\nonumber\\ %
&& %
\times
\avelBb \e^{ 
\frac{i}{\hbar}\vq \cdot\left[ \vR  \left( s^{\prime}\right) 
-\vR  \left(  s\right)
\right] 
}
2i\sin\left[ \frac{1}{2\hbar}\vq \cdot\Delta \rv
\left( s^{\prime}\right)  \right] \sum_{\mu=\pm}\mu\e^{
-\frac{i}{\hbar}\mu\frac{1}{2}\vq \cdot\Delta\rv \left( s\right)
}
\frac{\partial}{\partial\vP }w\left[ \vP \left(
s\right)  +\mu\frac{1}{2}\Delta\vp \left(  s\right)  -\mu\vq %
\right]  \averBb ,
\end{eqnarray}
where in the first term it is used that the average momentum
$\avelb \vP \averb $ is
the Fermi momentum
$\vp _{\mathrm{F}}=p_{\mathrm{F}}v_{\mathrm{F}}^{-1}\vv %
_{\mathrm{F}}$ (see appendix \ref{App LongCorr}). At least
for a non-interacting system $\Delta\vp \left(  s\right)  $ is of
the order of the typical wave-vector of the external force field.
Our interest is the semi-classical limit so we will assume
$\Delta\vp \left(  s\right) \cdot\vp
_{\mathrm{F}}<k_{\mathrm{B}}T$ and neglect its contribution, i.e.\ we put
$\Delta\vp \left(  s\right)  =0$. The average velocity $\vv %
_{\mathrm{F}}$ is then
\begin{eqnarray}
&& \vv _{\mathrm{F}}=
\frac{1}{m}\vp _{\mathrm{F}}-\int_{s}%
^{t}ds^{\prime}\int\frac{d\vq }{\left(  2\pi\hbar\right) ^{2}}\int
\frac{d\omega}{2\pi}\,v_{\mathrm{F}}m^{-1}p_{\mathrm{F}}D_{\perp\perp}\left(
\vq ,\omega \right) 
\e^ {
\frac{i}{\hbar}\vq \cdot\left[ \vR
\left( s^{\prime}\right) -\vR  \left(  s\right)  \right] 
}
\e^ {
-i\omega\left( s^{\prime}-s\right)  
}
\nonumber\\
&&
\times
\avelBb 2\sin\left[  \frac{1}{2\hbar}\vq \cdot\Delta\rv %
\left(  s^{\prime}\right)  \right] 
\left\{  \sin\left[ \frac{1}{2\hbar}\vq \cdot
\Delta\rv \left(  s\right) \right] \sum_{\mu=\pm}\frac{\partial
}{\partial\vP }w\left[  \vP \left(  s\right)  -\mu\vq %
\right]  +i\cos\left[  \frac{1}{2\hbar}\vq \cdot\Delta\rv \left(
s\right)  \right]  \sum_{\mu=\pm}\mu\frac{\partial}{\partial\vP %
}w\left[  \vP \left(  s\right)  -\mu\vq \right]  \right\}
\averBb.\!
\nonumber\\
\end{eqnarray}
We are mainly interested in the long wave-length limit, so we will
neglect terms of order $q^{2}$ compared to terms of order $q$. We
also assume that both $\left|  \vP \left(  s\right)  \right| $ and
the corresponding velocity $\left|  \dot{\vR}  \left( s\right)
\right|  $ are distributed evenly around the Fermi surface. In this
approximation the second term inside the parenthesis can be
neglected. We assume the relative variation in the velocity around
the Fermi velocity $\vv _{\mathrm{F}}$ is small, i.e.\ the
temperature is much smaller than Fermi energy. It is thus possible
to approximate the path locally by a straight line i.e.\ $\vR
\left( s^{\prime}\right) -\vR  \left(  s\right) \approx \vv %
_{\mathrm{F}}\left(  s^{\prime}-s\right)   $. In the long
wave-length limit the transverse part of $\vq \cdot\Delta\rv $ is
bigger than the longitudinal part. The averaging procedure is
approximated by independently carrying out the longitudinal and
the transverse average. In the longitudinal average we use the
time-local approximation outlined in appendix \ref{App LongCorr},
so the longitudinal average of $\sum_{\mu=\pm}-\frac{\partial
}{\partial\vP }w\left[  \vP \left(  s\right)  -\mu\vq %
\right]  $ is approximately
\begin{equation}
\avelBb \sum_{\mu=\pm}-\frac{\partial}{\partial\vP %
}w\left[  \vP \left(  s\right)  -\mu\vq \right] \averBb %
\approx %
\vv _{\mathrm{F}}\int_{-\infty}^{\infty
}d\varepsilon\,\frac{\partial}{\partial\varepsilon}f_{\mathrm{F}}\left(
\varepsilon\right)
\sum_{\mu=\pm}\frac{\partial}{\partial\varepsilon
}f_{\mathrm{F}}\left(  \varepsilon-\mu\vv _{\mathrm{F}}\cdot \vq
\right)  ,
\end{equation}
where we have used $\vv _{\mathrm{F}}=\frac{\partial\varepsilon
_{\vP }}{\partial\vP }$. In the transverse average we approximate
\begin{equation}
2\sin\left[  \frac{1}{2\hbar}\vq \cdot\Delta\rv \left( s^{\prime
}\right)  \right]  \sin\left[ \frac{1}{2\hbar}\vq \cdot\Delta
\rv \left(  s\right)  \right]    %
\approx
\left\{  1-\cos\left[ \frac{1}{\hbar}\vq\cdot\Delta\rv \left(
s\right)
\right] \right\}  
\approx %
\left[  1-\exp\left(  -\frac{1}{2}q_{\mathrm{c}\perp}^{-2}%
q^{2}\right)  \right]
\end{equation}
with the result that the effective mass $m^{\ast}$, defined as $\vp %
_{\mathrm{F}}=m^{\ast}\vv _{\mathrm{F}}$, is given by
\begin{eqnarray}
\frac{1}{m^{\ast}} %
& \approx& %
\frac{1}{m}\left\{  1+2v_{\mathrm{F}}^{2}%
\int_{-\infty}^{\infty}ds^{\prime}\int\frac{d\vq }{\left( 2\pi
\hbar\right) ^{2}}\int\frac{d\omega}{2\pi}\,\Re D_{\perp\perp
}\left(  \vq ,\omega\right)  
\left[  1-\e^{
-\frac{1}{2}q_{\mathrm{c}\perp }^{-2}q^{2}}  \right]
\right.  
\nonumber\\ %
&& %
\left.
\times
\e^{\frac{i}{\hbar}\vq \cdot\vv _{\mathrm{F}}\left(
s^{\prime}-s\right)  }  
\e^{  -i\omega\left(
s^{\prime }-s\right)  }  
\int_{-\infty}^{\infty}d\varepsilon\,\frac{\partial}{\partial
\varepsilon}f_{\mathrm{F}}\left(  \varepsilon\right)  \frac{\partial}%
{\partial\varepsilon}f_{\mathrm{F}}\left(  \varepsilon-\vv _{\mathrm{F}%
}\cdot\vq \right)  \right\}
.\label{Eq: EffectiveVelocityMomentumFinalFormel}%
\end{eqnarray}
Apart from the extra factor of $\left[  1-\exp\left(  -\frac{1}{2}%
q_{\mathrm{c}\perp}^{-2}q^{2}\right)  \right]  $ 
this relation is identical to the usual relation derived from the self-energy.
Thus the contribution from
the short wave-length modes of the interaction to the renormalization of the
mass is identical to the standard result, while the long wave-length modes do
not contribute. The cross-over is at the width $\hbar q_{\mathrm{c}\perp}%
^{-1}$ and this is why we above have alluded to the width as an 
effective cut-off.

When the long-wavelength expression for the propagator $D_{\perp\perp}%
$,\ Eq.\ (\ref{Eq: GaugePropTransversePropInfrared}) in the case of Coulomb
interaction, is inserted into Eq.\ (\ref{Eq:
EffectiveVelocityMomentumFinalFormel}) the integral is ultraviolet divergent.
We use $p_{\mathrm{F}}$ as an approximative upper cut-off and get the
effective mass%
\begin{equation}
m^{\ast}\approx m+m \frac{4}{\pi^{3/2}}
\frac{p_{\mathrm{F}}^{2}}{2 m}\frac{1}{\coulener}
\log\frac{p_{\mathrm{F}}^{2}}{2q_{\mathrm{c}\perp}^{2}}.
\end{equation}
For composite fermions the renormalization of the mass is quite strong so the
effective mass
\begin{equation}
m^{\ast}\approx
\frac{2}{\pi^{3/2}}
\frac{p_{\mathrm{F}}^{2}}{\coulener}
\log\frac{p_{\mathrm{F}}^{2}}{2q_{\mathrm{c}\perp}^{2}}%
\end{equation}
is approximately independent of the bare mass $m$.

The prefactor
\begin{equation} 
m \frac{4}{\pi^{3/2}}
\frac{p_{\mathrm{F}}^{2}}{2 m}\frac{1}{\coulener}
\approx  1.9 \: m ,
\end{equation}
where
$\coulener=\frac{e^2}{4\pi\permittivity}\sqrt{n}$
is the average Coulomb energy and we as a typical values
\cite{WillettEksperiment} 
have used 
$n= 1.6 \times 10^{15} \mathrm{m}^{-2}$, 
$\permittivity =13 \: 
\permittivity_0 
$, and 
$m\approx 0.067 \:m_0
$, where $\permittivity_0$ and $m_0$ are the vacuum values of 
the permittivity and the electron mass respectively.

The argument of the logarithm is given by %
Eq.\ (\ref{Eq SemiClasSep Final general})
so self-consistently the effective mass is
\begin{equation}
\frac{m^{\ast}}{m}\approx 1+\frac{4}{\pi^{3/2}}
\frac{p_{\mathrm{F}}^{2}}{2 m}
\frac{1}{\coulener}
\log\left[
\frac{1}{2}+\frac{\pi^{9/4}}{18}
\sqrt{\frac{\coulener}{\T}}
\left(\frac{4}{\pi^{3/2}}
\frac{p_{\mathrm{F}}^{2}}{2 m}\frac{1}{\coulener}\right)^{2}
\left(\frac{m}{m^{\ast}}\right)^2
\right]
\approx
6.3 ,
\label{Eq EffMass Numberestinate}
\end{equation}
where the estimate is valid at a temperature of $\sim 0.13$~K. 
The cut-off at this temperature happens at distances 
$\hbar q_{\mathrm{c}\perp}^{-1}\approx 6 \lambda_{\mathrm{F}}
\approx 40$~nm.  

As expected the semi-classical effective mass 
Eq.\ (\ref{Eq EffMass Numberestinate}) is 
smaller than the mass deduced from the 
activation energy and Shubnikov-de Haas measurements 
\cite{WillettEksperiment}, actually about half as big.
In the experiment by Willett et al. 
\cite{WillettEksperiment} the estimate 
Eq.\ (\ref{Eq EffMass Numberestinate}) leads to a 
cyclotron frequency that
at the
secondary resonance position is about twice as big as the used SAW frequency
$10.7$~GHz. A naive argument \cite{WillettEksperiment} leads to 
the expectation
that the secondary resonance can only be observed if the cyclotron frequency 
is much bigger than the SAW frequency. It is thus a matter of taste whether
an `apparent inconsistency' between the geometric resonance 
experiments and the size
of the effective mass arises. We here want to stress the word `apparent' 
since Mirlin and W\"{o}lfle \cite{Wølfe} has shown that the 
secondary resonance can be observed even when the cyclotron frequency  
equals the SAW frequency.  


%

\section{Conclusion}

In this paper we have developed a single-particle path integral
description of transport for composite fermions, thereby extending 
the Caldeira-Leggett formalism for
a single-particle interacting with an environment. Here the
environment is the Chern-Simons gauge field. Our approach is modeled 
over the work by Golubev and Zaikin on the
electron gas without a magnetic field. As in the
Caldeira-Leggett formalism there is not one, but two paths 
in the path integral, one propagating
forward in time and one backward. A new feature compared to the 
Caldeira-Leggett formalism is that the distribution
function enters into the real part of the single-particle action. 
The imaginary part of the action is 
determined by the fluctuation spectrum, just as in the 
Caldeira-Leggett formalism.

In a perturbative description of the CF's
the long wave-length magnetic field fluctuations give rise to an
infrared divergence of the self-energy and the vertex corrections,
separately. In a transport measurement these divergences are 
known to cancel. 
In our formalism the infrared divergence does not appear because the vertex 
and self-energy contributions are never split. Not only is the infrared 
divergence absent, but the renormalization in the infrared limit is negligible.
This result is obtained in a natural way by working in a real space 
formulation, 
since the magnetic interaction acts via fluxes, i.e.\  
areas. In the path integral formalism the size of the particle is defined
as the separation between the forward and backward propagating paths. Only
the magnetic field fluctuations with a wave-length shorter than 
this particle size contributes noticeably to the renormalization of the mass. 
In GaAs at $T\sim 0.13$~K we find the mass renormalization
due to  the magnetic field fluctuations to be roughly 6 times the bare mass. 
Furthermore we have shown that in
the semi-classical limit the separation between the two paths, i.e.\
 the 
size of the particle,
scale as $T^{-1/4}|\log T|^{-1}$. This implies 
the effective 
cut-off in frequency scale
as roughly $T ^{1/2}|\log T|^2$. As usual the mass scales approximately with
the logarithm of the frequency cut-off. 

The linear response regime around equilibrium is 
the major focus of our work, but the formalism is
still valid out of equilibrium. In the non-interacting limit, 
as shown in Fig.\ \ref{Fig3kubo}, 
the single-particle path integral
technique possesses the feature that the effect of the non-equilibrium 
distribution function is gathered completely within
the stimulus, while the propagation is identical to the propagation 
in equilibrium. This feature is shared with other 
out of equilibrium techniques i.e.\ the  Landauer-B\"{u}ttiker description
of transport through a non-interacting  part of the system and the 
Boltzmann equation. In fact the classical limit of 
the single-particle path integral formalism leads very naturally to 
the Boltzmann equation.  
The single-particle path integral technique is thus a promising technique 
for the description of interacting systems both in and out of equilibrium, 
especially it may prove useful for studying 
systems out of equilibrium not easily treated by other methods.

\section{Acknowledgement}

K.A.E.\ and H.B.\ both acknowledge support from the Danish Natural
Science Research Council through Ole R{\o}mer Grant no.\ 9600548.

\appendix


\section{Equations of motion} \label{AppDiff}

In this appendix we are going to derive differential
equations for the composite fermion contribution to the effective
action $\Seffa[a;t]$ and the density matrix $\rhoac(\A-a)$ in the
presence of the Chern-Simons field $\A-a$  by
differentiating the composite fermion part of Eqs.\ (\ref{Eq:
DefSeffa}) and (\ref{Eq: Defrhoac1}).

In order to differentiate quantities in a fixed gauge field
background at time $s$ it is necessary to let the Chern-Simons
field and the Grassmann fields $\bar{\psi }$ and $\psi$ be defined
on $\Ct$, where $t>s$. The effective action for both the
Chern-Simons field and the Grassmann variables defined in
Eq.\ (\ref{action1CompositeFermion}) is then
\begin{equation}
\Seff[\bar{ \psi}, \psi, a; s] = \SCS[a;t]
-
e \int _{\Ct} d \sigma^{\prime} \int d \rv  a_0(\rv, \sigma
^{\prime}) n + \Scom[\bar{ \psi}, \psi, a; s;t] ,
\label{Eq: Appaction1eff}
\end{equation}
where
\begin{equation}
\Scom[\bar{ \psi}, \psi, a; s;t]=
\int _{\Ct} d \sigma^{\prime} \int d
\rv \bar{\psi } (\rv, \sigma ^{\prime} ) i \hbar \frac{\partial
}{\partial \sigma ^{\prime} } \psi(\rv, \sigma ^{\prime}) - \int
_{\Cs} d \sigma^{\prime} \Hn[\A - a].
\label{Eq: Appaction1CompositeFermion}
\end{equation}
The second term in Eq.\ (\ref{Eq: Appaction1eff}) can be gauged away for
$t>s^{\prime}>s$, since
there is no Hamiltonian. The terms that depend on the composite
fermion Grassmann variables are gathered in $\Scom[\bar{ \psi}, \psi, a; s;t]$
and it
is its contribution we are going to consider in this appendix.

Notice the effective action $\Seff$ in the definition of $\rhoac$, %
Eq.\ (\ref{Eq: Defrhoac1}), can be replaced by $\Scom$, since the
difference cancels between the numerator and denominator in %
Eq.\ (\ref{Eq: Defrhoac1}).

\subsection{The effective action}
\label{AppSec DiffEffectiveAction}

The composite fermion contribution to the effective action
$\Seffa[a;s]$ is
\begin{equation}
\Seffa[a;s] - \SCS[a;t] + e \int _{-\infty}^{s}
d \sp \int d \rv  \Delta a_0(\rv, \sp ) n = \frac{\hbar}{i}
\log
\gbl 1 \gbr ,
\label{Eq: AppDiffEffActionCompositDel}
\end{equation}
where $\Delta a_0(\sp) =a_0(\sp,\fin)-a_0(\sp,bi)$ is the difference
field and the short-hand notation
\begin{equation}
\gbl X \gbr = \int \cd \bar{\psi }  \int \cd \psi \exp
\left( \frac{i}{\hbar} \Scom[\bar{ \psi}, \psi, a; s;t]
\right) \, X
\label{Def gbl gbr shorthand}
\end{equation}
for the Grassmann integral has been introduced. %
The derivative  of Eq.\ (\ref{Eq: AppDiffEffActionCompositDel}) is
\begin{equation}
\pderives \big\{
\text{Eq.\ (\ref{Eq: AppDiffEffActionCompositDel})} 
\big\}= -
\gbl 1 \gbr ^{-1}
\gbl \Hn[\A(s) - a(s,\fin)]-\Hn[\A(s) - a(s,\bi)] \gbr .
\end{equation}
Per definition of the path integral is the right-hand side equal to 
the expectation value of $ -\Big\{\Hn[\A(s) - a(s,\fin)]$
$-\Hn[\A(s) - a(s,\bi)]\Big\}$
in the presence of the contour gauge field $\A-a$. It can be
expressed in terms of the density matrix as
\begin{equation}
\pderives \text{Eq.\ (\ref{Eq: AppDiffEffActionCompositDel})} =
-\int d \rv \left(
e \Delta a_0(\rv, s) \rhoac(\A-a; \rv, \rv; s ) + \vda(\rv, s)
\cdot \vja(\A-a; \rv, s) \right ) ,
\end{equation}
where the current in the presence of the gauge field $\A-a$: $\vja(\A-a; \rv,
s)$ is defined in Eq.\ (\ref{Eq: CurrentPresenceGaugeFieldHistory}).
Eq.\ (\ref{Eq: EqSeffa Generel}) then follows upon integration
using the conventional assumption of adiabatic turning on of the
gauge field.

\subsection{The density matrix}

The density matrix in the presence of the Chern-Simons gauge field
$\rhoac$ is defined in Eq.\ (\ref{Eq: Defrhoac1}). At a time $s<t$ the action
$\Seff[\bar{ \psi}, \psi, a;s]$ can be replaced by $\Scom[\bar{ \psi}, \psi, a; s;t]$.
In terms of the above %
short-hand notation
\begin{equation}
\rhoac\left(\rv,\rv^{\prime},s\right) = \gbl 1
\gbr^{-1}
\gbl \bar{\psi } ( \rv ^\prime, s ) \psi(\rv, s) \gbr.
\end{equation}
When $s$ is varied both the
numerator and the denominator varies:
\begin{equation}
i \hbar \pderives \rhoac\left(\rv,\rv^{\prime},s\right) %
= \gbl 1
\gbr^{-1}
i \hbar \pderives \gbl \bar{\psi } ( \rv ^\prime, s ) \psi(\rv, s) \gbr%
-
\gbl 1\gbr^{-1}
\gbl \bar{\psi } ( \rv ^\prime, s ) \psi(\rv, s) \gbr
i \hbar \pderives \gbl 1 \gbr %
.
\label{Eq: AppCherndiff formalderivative rho}
\end{equation}
In the above subsection \ref{AppSec DiffEffectiveAction} %
the derivative of the denominator 
$i \hbar \pderives \gbl 1 \gbr $
was calculated. The contribution from the first term (the numerator) is
\begin{equation}
 \gbl 1
\gbr^{-1} i \hbar \pderives \gbl \bar{\psi } ( \rv ^\prime, s ) %
\psi(\rv, s) \gbr  =
\end{equation}
\begin{equation}
\gbl 1 \gbr^{-2} 
\Big\{
\gbl \Hn[\A(s) - a(s,\fin)]\bar{\psi } ( \rv ^\prime, s ) \psi(\rv, s)
-\bar{\psi } ( \rv ^\prime, s ) \psi(\rv, s)\Hn[\A(s) - a(s,\bi)] \gbr 
\Big\} .
\label{Eq:AppDiffNum1}
\end{equation}
We here used
\begin{equation}
\gbl \bar{\psi } ( \rv ^\prime, s+\Delta s ) \psi(\rv, s+ \Delta
s)\gbr
=
\gbl \bar{\psi } ( \rv ^\prime, s) \psi(\rv, s)\gbr ,
\end{equation}
since without a Hamiltonian the system does not evolve. The action
$\Scom$ is quadratic in the Grassmann variables $\bar{\psi}$ and
$\psi$, so Eq.\ (\ref{Eq:AppDiffNum1}) can be evaluated with
the help of Wick's theorem. The Hartree disconnected diagram like
pairings exactly cancels the contribution from the variation of
the denominator in Eq.\ (\ref{Eq: AppCherndiff formalderivative rho}), 
$
\gbl 1 \gbr ^{-1}%
\gbl \bar{\psi } ( \rv ^\prime, s ) \psi(\rv, s) \gbr
i \hbar \pderives \gbl 1 \gbr $, so paying due attention to 
the contour ordering
\begin{eqnarray}
i \hbar \pderivet \rhoac(\rv,\rv^{\prime},t) &=&
\int d \rv ^{\prime \prime} \int d \rv^{\prime \prime \prime}
\left(\delta(\rv-\rv^{\prime \prime})-\rhoac[\rv,\rv^{\prime
\prime};t]\right)
\Ham(\rv^{\prime \prime}, \rv^{\prime \prime \prime};t,\fin)
\rhoac[\rv^{\prime \prime \prime},\rv^{\prime};t]
\nonumber\\
&&
-
\int d \rv ^{\prime \prime} \int d \rv^{\prime \prime \prime}
\rhoac[\rv,\rv^{\prime \prime};t]
\Ham(\rv^{\prime \prime}, \rv^{\prime \prime \prime};t,\bi)
\left( \delta(\rv^{\prime \prime \prime}-\rv^{\prime})-
\rhoac[\rv^{\prime \prime \prime},\rv^{\prime};t]\right)
.
\label{EqApp Diff rhoac}
\end{eqnarray}
This is Eq.\ (\ref{Diff rhoac}) written out.


\section{Linearization of the density matrix differential
equation} %
\label{Sec: appDriveD listed}

The stimulus in the linearized equation of motion for $\delta
\rhoac(\A-a; \rv, \rv ^\prime, t) $, Eq.\ %
(\ref{LinearDensityMatrixDiff}), $\D = \D_1 + \D_2$ is split into
two parts. The explicit form for $\D_1$, which is independent of
$\Delta \va$, is
\begin{eqnarray} i \hbar \D_1(\rv, \rv ^\prime ,t) &=&
\left[-\frac{\partial}{\partial\rv}\int_{\rv^{\prime}}^{\rv}d\vs
\cdot \vAt(\vs) + \vAt(\rv)\right]\cdot
\frac{1}{m}\left[\frac{\hbar}{i}\frac{\partial}{\partial \rv}+e
\A(\rv) - e \bar{\va}(\rv)\right]\rhoac(\rv,\rv^{\prime})
\nonumber\\ && + \left[-\frac{\partial}{\partial
\rv^{\prime}}\int_{\rv^{\prime}}^{\rv}d\vs \cdot \vAt(\vs) -
\vAt(\rv^{\prime})\right]\cdot
\frac{1}{m}\left[-\frac{\hbar}{i}\frac{\partial}{\partial
\rv^{\prime}}+ e \A(\rv^{\prime}) - e
\bar{\va}(\rv^{\prime})\right] \rhoac(\rv,\rv^{\prime})
\nonumber\\ && +\rhoac(\rv,\rv^{\prime})\frac{\hbar}{2 i m}
\frac{\partial}{\partial \rv} \left[-\frac{\partial}{\partial
\rv}\int_{\rv^{\prime}}^{\rv}d\vs \cdot \vAt(\vs) -
\vAt(\rv)\right] \nonumber\\ &&
-\rhoac(\rv,\rv^{\prime})\frac{\hbar}{2 i m}
\frac{\partial}{\partial \rv^{\prime}}
\left[-\frac{\partial}{\partial
\rv^{\prime}}\int_{\rv^{\prime}}^{\rv}d\vs \cdot \vAt(\vs) -
\vAt(\rv^{\prime})\right] \nonumber\\ &&
-\int_{\rv^{\prime}}^{\rv}d \vs \cdot \left(\pderivet
\vAt+\frac{\partial \phit}{\partial \vs} \right)
\rhoac(\rv,\rv^{\prime}).
\end{eqnarray}
The explicit form of $\D_2$, which contains all the dependence
upon $\Delta \va$, is %
\begin{eqnarray} %
i \hbar \D_2(\rv, \rv ^\prime ,t) &=&
\int d \rv^{\prime \prime}(1-\rhoac)(\rv,\rv^{\prime \prime})
\frac{-e \Delta \va(\rv^{\prime \prime})}{2m}\cdot
\left[-\frac{\partial}{\partial \rv^{\prime
\prime}}\int_{\rv^{\prime}}^{\rv^{\prime \prime}} d\rv^{\prime
\prime \prime} \cdot \vAt(\rv^{\prime \prime
\prime})+\vAt(\rv^{\prime \prime})\right] \rhoac(\rv^{\prime
\prime},\rv^{\prime}) \nonumber\\ 
&& \ekstra{\hspace{-2 cm}}
+ \int d \rv^{\prime \prime}
\rhoac(\rv,\rv^{\prime \prime}) \frac{-e \Delta \va(\rv ^{\prime
\prime})}{2m}\cdot \left[\frac{\partial}{\partial \rv^{\prime
\prime}}\int^{\rv}_{\rv^{\prime \prime}} d\rv^{\prime \prime
\prime} \cdot \vAt(\rv^{\prime \prime \prime})+\vAt(\rv^{\prime
\prime})\right] (1-\rhoac)(\rv^{\prime \prime},\rv^{\prime})
\nonumber\\ 
&& \ekstra{\hspace{-2 cm}}
- \frac{i}{2 \hbar} \int d
\rv^{\prime \prime} \rhoac(\rv,\rv^{\prime \prime})
\left[\int_{\rv^{\prime}}^{\rv} d\rv^{\prime \prime \prime} \cdot
\vAt(\rv^{\prime \prime \prime}) -\int_{\rv^{\prime}}^{\rv^{\prime
\prime}} d\rv^{\prime \prime \prime} \cdot \vAt(\rv^{\prime \prime
\prime}) -\int_{\rv^{\prime}}^{\rv^{\prime \prime}} d\rv^{\prime
\prime \prime} \cdot \vAt(\rv^{\prime \prime \prime})\right]
\nonumber\\ 
&& 
\times
\frac{-e \Delta \va}{2m}\cdot
\left[\frac{1}{2 m}\Delta \va \left[\frac{i}{\hbar}\frac{\partial}
{\partial \rv^{\prime \prime}} +\vA(\rv^{\prime \prime})\right] -
\Delta a_0\right] \rhoac(\rv^{\prime \prime},\rv^{\prime})
\nonumber\\ 
&& \ekstra{\hspace{-2 cm}}
-
\frac{i}{2 \hbar} \int d \rv^{\prime \prime} \rhoac(\rv^{\prime
\prime},\rv^{\prime}) \left[\int_{\rv^{\prime}}^{\rv} d\rv^{\prime
\prime \prime} \cdot \vAt(\rv^{\prime \prime \prime})
-\int_{\rv^{\prime}}^{\rv^{\prime \prime}} d\rv^{\prime \prime
\prime} \cdot \vAt(\rv^{\prime \prime \prime})
-\int_{\rv^{\prime}}^{\rv^{\prime \prime}} d\rv^{\prime \prime
\prime} \cdot \vAt(\rv^{\prime \prime \prime})\right]
\nonumber\\ 
&& 
\times
\frac{-e \Delta \va}{2m}\cdot \left[\frac{1}{2 m}
\Delta \va \left[-\frac{i}{\hbar}\frac{\partial} {\partial
\rv^{\prime \prime}} +\vA(\rv^{\prime \prime}) \right] - \Delta
a_0\right] \rhoac(\rv,\rv^{\prime \prime}) . %
\label{DrivRhoDef}
\end{eqnarray}


\section{The longitudinal separation}

\label{App LongCorr}

In this appendix we are interested in the distribution of the
longitudinal separation $\frac{1}{P}\vP \cdot \Delta \rv $ and its
conjugate variable the size of the momentum $P$. For simplicity we
will only consider the case of a scalar interaction.

The first step is to rewrite the free particle action such that
the longitudinal and transverse directions appear explicitly.

To this end we use polar coordinates for the momentum
\begin{equation}
\vp =\left( \matrix{ %
p\cos\frac{p_{\eta}}{p}
\cr
p\sin\frac{p_{\eta}}{p}%
}
\right),
\end{equation}
$\frac{p_{\eta}}{p}$ is the angle with respect to an arbitrarily
defined $x$-axis. $p_{\eta}$, loosely speaking the arc-length along
the Fermi surface, has been chosen as our variable instead of the
angle $\frac{p_{\eta}}{p}$ in order to avoid a complicated
Jacobian. In the scalar approximation the interaction is
independent of $p_{\eta}$, which therefore can be integrated out.
The free particle action on the forward part of the contour is
\begin{equation}
\int_{0}^{t}ds\,\left(\dot{\rv}\cdot
\vp-\frac{\vp ^{2}}{2m}\right) .%
\end{equation}
The stationary phase approximation for $p_{\eta}$ reads
\begin{equation}
p_{\eta}=p\arctan\frac{\dot{y}}{\dot{x}} .%
\label{Eq: LowFreq Statio Phase Ang Momentum}
\end{equation}
There are two solutions to this equation corresponding to the
momentum and the velocity being parallel or anti-parallel. Here we
only keep the parallel solution. The resulting stationary phase
action on the forward contour is
\[
\int_{0}^{t}ds\,\left(p v - \frac{\vp ^{2}}{2m}\right),%
\]
where $v=\sqrt{\dot{x}^{2}+\dot{y}^{2}}$ is the velocity. A
similar result holds on the backward path, but with the opposite
overall sign. We are interested in discussing the separation
between the two paths, so we change representation from the
position and momentum along each path individually to the average
$\vR  \left(  s\right)  =\frac{1}{2}
\left[\rv %
\left(  s,+\right)  +\rv \left(  s,-\right) \right]$ and the
difference $\Delta\rv \left(  s\right)  =\rv \left( s,+\right)
-\rv \left(  s,-\right)  .$ In a translation invariant system one
has the choice between using $\vR  $ or $\dot{\vR}$ as the
variables in the path integral. Here we choose $\dot{\vR}$ and we
parameterize it using the velocity
$V(s)=\left|\dot{\vR}(s)\right|$ and the polar angle
$\frac{V_{\phi}\left(  s\right)  }{V\left(  s\right) }$
\begin{equation}
\dot{\vR} \left(  s\right)  = \left( \matrix{
V\left( s\right)  \cos\frac{V_{\phi}\left(  s\right)  }{V\left(
s\right) }
\cr
V\left(  s\right)  \sin\frac{V_{\phi}\left(  s\right)  }{V\left(  s\right)  }%
}
\right)  =V\left(  s\right)  \bn  \left(  s\right) .
\end{equation}
The unit vector $\bn  \left(  s\right)  $ points along the
direction of the average propagation. $\bn  \left( s\right) $ can
also be thought of as a unit vector on the
Fermi surface, since above in Eq.\ %
(\ref{Eq: LowFreq Statio Phase Ang Momentum}) we only used the
stationary phase condition where the momentum is aligned with the
velocity.

Neglecting third order terms in the difference variables the free particle
action is
\begin{eqnarray}
&&
\sum_{\mu}\int_{0}^{t}ds\, \left[p\left(  s,\mu\right)  v\left(
s,\mu\right) - \frac{\vp \left(  s,\mu\right)
^{2}}{2m}  \right] %
\approx  %
\nonumber\\
&&
\qquad
\int _{0}^{t}ds\, \left[P\left(  s\right) \bn  \left( s\right)
\cdot \Delta\dot{\rv}\left( s\right)  +\Delta
p\left( s\right)  V\left( s\right) \right]
-\int _{0}^{t}ds\, \frac{P\left( s\right) \Delta p\left(  s\right)
}{m} 
+\blamk  \cdot \left[\left(  \vR  \left(  t\right) -\vR  \left(
0\right)  \right)  -\int_{0}^{t}ds\,V\left( s\right)  \bn  \left(
s\right)  \right] ,%
\end{eqnarray}
where $\blamk $ is a Lagrange multiplier vector that has been
introduced in order to deal with the boundary condition in the
transition from $\vR  $ to $\dot{\vR}$. $\frac{1}{\hbar}|\blamk| V
\tau_{\text{tr}}$ is a small number, when the feeding-in and the
pick-up are not local in space. As an illustration assume that the
initial and final position is distributed according to isotropic
Gaussians centered around $\vR _{\mathrm{i}}^{0}$ and $\vR
_{\mathrm{f}}^{0}$ respectively. The sum of the variances is
denoted by $\left( \frac{\hbar}{\Delta k}\right) ^{2}$. The term
$\blamk \cdot\left(  \vR  _{\mathrm{f}}-\vR  _{\mathrm{i}%
}\right)  $ in the action is then replaced by
\begin{equation}
\blamk \cdot\left(  \vR  _{\mathrm{f}}^{0}-\vR  %
_{\mathrm{i}}^{0}\right)
+\frac{i\hbar}{2}\frac{k^{2}}{\Delta
k^{2}}%
\end{equation}
after the integration with respect to $\vR  _{\mathrm{i}}$ and
$\vR  _{\mathrm{f}}$ has been performed. The typical size of
$\blamk $ is now given by the smallest of the numbers $\frac{\hbar
}{\left| \vR  _{\mathrm{f}}-\vR  _{\mathrm{i}}\right| }$ and
$\Delta\lambda.$ The last number is in the semi-classical limit
small. In the case of a periodic driving $\blamk$ is the
wave-vector of the external field. Anyway in the semi-classical
limit we will neglect the contribution from $\blamk$.

\subsection{The interaction part: general formulation}

We first Taylor expand the interaction part of the action $\Sint$,
the scalar part of Eq.\ %
(\ref{Golchern interaction action full}), to second order in the
difference variable $\vdr$. Furthermore, we use a time local
approximation. That is we approximate the quadratic terms $\vdr(s)
\vdr(s^{\prime})$ as $\vdr(s)^2$. Using this
\begin{equation}
\Sint \approx \frac{i}{2}\int_{0}^{t}ds\Delta\rv\left( s\right)
\cdot {\mathbf{F}}\left[  \vR  ,s\right] \cdot\Delta \rv \left(
s\right) -\int_{0}^{t}ds \Delta \rv \left( s\right)
\cdot{\mathbf{h}}\left[ \vR  ,P,s\right] ,
\end{equation}
where%
\begin{eqnarray}
{\mathbf{h}}\left(  \vR  ,P,s\right)    & = & h\left( V\left(
s\right) ,P,s\right)  \bn  \left(  s\right)  \\ {\mathbf{F}}\left(
\vR  ,s\right)  _{\alpha\beta}  & = & \left(
\delta_{\alpha\beta}-\bn  _{\alpha}\bn  _{\beta}\right)
F_{\mathrm{T}}\left[  V\left(  s\right)  \right] +\bn  _{\alpha
}\bn  _{\beta}F_{\mathrm{L}}\left[ V\left(  s\right)  \right]
\end{eqnarray}
with
\begin{eqnarray}
h\left(  V\left(  s\right)  ,P,s\right)    & = & %
\int_{0}^{t}ds^{\prime\prime} \frac{\partial
D_{00}}{\partial\left| \vR  \right|  }\left(  V\left( s\right)
\left( s-s^{\prime\prime}\right)  ,s-s^{\prime\prime}\right)
\sum_{\mu^{\prime\prime}}\left[ \frac{1}{2}-w\left( P\left(
s^{\prime\prime+}\right)  +\left(  -1\right)  ^{\mu^{\prime\prime}}%
\frac{\Delta p\left(  s^{\prime\prime+}\right)  }{2}\right)  \right]
\\
F_{\mathrm{L}}\left[  V\left(  s\right)  \right]  &=& %
\int_{0}^{t}ds^{\prime }\frac{\partial^{2}C_{00}}{\partial\left|
\vR  \right|  ^{2}}\left( V\left(  s\right)  \left|
s^{\prime}-s\right|  ,s^{\prime}-s\right)
\\ %
F_{\mathrm{T}}\left[  V\left(  s\right)  \right]
&= & %
\int_{0}^{t}ds^{\prime }\frac{1}{V\left(  s\right)  \left|
s^{\prime}-s\right|  }\frac{\partial C_{00}}{\partial\left| \vR
\right|  }\left(  V\left(  s\right)  \left|
s^{\prime}-s\right|  ,s^{\prime}-s\right) . %
\label{Eq CT definition}
\end{eqnarray}
As a further simplification we have above also approximated the
average path by a straight line inside the kernels. In fact we will below treat both $ F_{\mathrm{T}}$ and $ F_{\mathrm{L}}$ as constants.


In this approximation the action separate into a longitudinal and
a transverse part. We define
\begin{equation}
\vdr\left(  s\right)  =\drl\left(  s\right) \bn  \left( s\right)
+\drt \left(  s\right) \zhat \times \bn  \left( s\right).
\end{equation}
In the Wigner variables basis of the propagator we use a polar
coordinate representation of the initial and final momenta
\begin{eqnarray}
\vP _{\mathrm{i}}  & = & P_{\mathrm{i}}\left( \matrix{
\cos\phi_{\mathrm{i}}
\cr
\sin\phi_{\mathrm{i}}%
}
\right)  \\ %
\vP _{\mathrm{f}}  & = & %
P_{\mathrm{f}}\left( \matrix{
\cos\phi_{\mathrm{f}}
\cr
\sin\phi_{\mathrm{f}}%
}
\right).
\end{eqnarray}
In the full propagator we now split out all the factors related to
the transverse direction into the so-called transverse propagator
\begin{equation}
J_{\text{T}}= 
\int\cd   V_{\phi}\int\cd   \drt\,%
\frac{2\pi\hbar}{P_{\mathrm{f}} }\delta\left[
\phi_{\mathrm{f}}-\frac{V_{\phi}\left(  t\right)
}{V\left( t\right)  }\right]%
\frac{2\pi\hbar}{P_{\mathrm{i}}}\delta\left[
\phi_{\mathrm{i}}-\frac{V_{\phi}\left(  0\right)  }{V\left(
0\right) }\right] 
\e ^{ 
-\frac{1}{\hbar}\int_{0}^{t}ds\;
\left[\frac{1}{2}F_{\text{T}} \drt\left(
s\right)^2 + i \drt(s) P(s)\frac{d}{d s} \frac{V_{\phi}(s)}{V(s)}
\right]
}
. %
\label{Eq Transverseprop Def}
\end{equation}
In terms of $J_{\text{T}}$ the full propagator is after
integrating out $\vdr(0)$ and $\vdr(t)$
\begin{eqnarray}
J  & =&\int\cd   P\int
\cd   \Delta p\int\cd   V\int\cd   V \drl \, %
2\pi\hbar\delta\left[  P_{\mathrm{f}%
}-P\left(  t\right)  \right]  2\pi\hbar\delta\left[
P_{\mathrm{i}}-P\left( 0\right)  \right] J_{\mathrm{T}}\nonumber\\ %
&&%
\times
\e^ { 
\frac{i}{\hbar}\int_{0}^{t} ds \left[\Delta
p\left( s\right) V\left( s\right) - \frac{1}{m}P\left( s\right)
\Delta
p\left(  s\right)  \right] 
- \frac{1}{\hbar} \int _{0}^{t} d s \,
\left[\frac{1}{2}\drl(s)^2 F_{\text{L}} + i
\drl(s)\left(\frac{d}{d s}P(s)+h[V(s), P, s] \right)\right] 
}.
\label{Eq: LowFre LongitudProp}
\end{eqnarray}
To lowest order the transverse propagator is independent of the dynamics of the longitudinal variables, so we will treat $ J_{\mathrm{T}}$ as a constant and absorb it in the overall normalization of the path integral. The study of the remaining longitudinal propagator is the topic of the next subsection.

\subsection{The interaction part: the longitudinal contribution}

In the propagator Eq.\
(\ref{Eq: LowFre LongitudProp}) we will neglect the dependence of $h$ upon
$\Delta p$. $V\left(  s\right)  $ thus equals $m^{-1}P\left(  s\right)  $. The
propagation described by Eq.\ %
(\ref{Eq: LowFre LongitudProp}) is then equivalent to the quantum propagation
problem
\begin{equation}
-\hbar\pderives =\frac{1}{2}\left(  \frac{\hbar}{i}%
\frac{\partial}{\partial P}\right)  ^{2}F_{\mathrm{L}}\left(  P\right)
-\hbar\frac{\partial}{\partial P}h\left(  P\right)
\label{Eq PropProbLongBeforeGauge}%
\end{equation}
We neglect the momentum dependence of $F_{\mathrm{L}}$. In equilibrium is
\begin{equation}
h\left(  P\right)  \approx\alpha\frac{1}{2}\tanh\left[  \frac{v_{\mathrm{F}}%
}{2k_{\mathrm{B}}T}\left(  P-p_{\mathrm{F}}\right)  \right]  ,
\end{equation}
where the constant  $\alpha$ in the high temperature limit is related to the
fluctuations through%
\begin{equation}
\frac{\hbar F_{\mathrm{L}}}{\alpha}=\frac{k_{\mathrm{B}}T}{v_{\mathrm{F}}}.
\end{equation}
The last term in Eq.\ %
(\ref{Eq PropProbLongBeforeGauge}) is gauged away by the transformation
$\phi\left(  P\right)  \left(-\frac{\partial
f_{\mathrm{F}}}{\partial p}\right)^{-1/2}$  and the
propagation problem Eq.\ %
(\ref{Eq PropProbLongBeforeGauge}) is equivalent to
\begin{equation}
-2\frac{\hbar F_{\mathrm{L}}}{\alpha^{2}}\pderives \phi = 
-\frac{\hbar^{2}F_{\mathrm{L}}^{2}}{\alpha^{2}}\frac{\partial^{2}\phi
}{\partial P^{2}} 
+\left[  -\frac{\hbar
F_{\mathrm{L}}}{\alpha}\frac{\partial  }{\partial P}\tanh\left[
\frac{1}{2}\frac{v_{\mathrm{F}}}{k_{\mathrm{B}}T}\left(  P-p_{\mathrm{F}%
}\right)  \right]
+\tanh^{2}\left[  \frac{1}{2}\frac
{v_{\mathrm{F}}}{k_{\mathrm{B}}T}\left(  P-p_{\mathrm{F}}\right)  \right]
\right]  \phi.\label{Eq PropProbLongAfterGaugeLocal}%
\end{equation}
The ground state of the quantum problem Eq.\ %
(\ref{Eq PropProbLongAfterGaugeLocal}) is the derivative of the Fermi
distribution function and the ground state energy is zero. The typical decay
time of the excited states is given by $\frac{\hbar F_{\mathrm{L}}}{\alpha
^{2}}$. This can be identified as the correlation time scale of $\Delta
r_{||}$ or $P$ with themselves. It is also the time scale it takes before the
boundary conditions cannot be felt anymore. For the Coulomb interaction this
time-scale is approximately $\hbar\left(  k_{\mathrm{B}}T\right)  ^{-1}\hbar
\gc p_{\mathrm{F}}^{2}\left(  v_{\mathrm{F}}p_{\mathrm{F}}\right)  ^{-1}$. The
ground-state wave-function of the original quantum propagation problem Eq.\ %
(\ref{Eq PropProbLongBeforeGauge}) is proportional to $\sqrt{-\frac{\partial
f_{\mathrm{F}}}{\partial p}}$. This implies the distribution of the
longitudinal momentum or energy is given by the derivative of the Fermi
distribution function, while the average of the width squared is
\begin{equation}
\left\{  \int_{-\infty}^{\infty}dp\,\left(  -\frac{\partial f_{\mathrm{F}}%
}{\partial p}\right)  \right\}  ^{-1}\int_{-\infty}^{\infty}dp\,\left(
\frac{\hbar}{2}\frac{\partial^{2}f_{\mathrm{F}}}{\partial p^{2}}\right)
^{2}\left(  -\frac{\partial f_{\mathrm{F}}}{\partial p}\right)  ^{-1}=\frac
{1}{3}\frac{\hbar^{2}v_{\mathrm{F}}^{2}}{\left(  k_{\mathrm{B}}T\right)  ^{2}%
}.
\end{equation}




\end{document}